\documentclass[12pt]{article}
\usepackage[dvips]{epsfig,graphics}
\usepackage{graphicx}
\usepackage{amsmath,amsfonts,amssymb}
\newcommand{\gsim}{\gtrsim}

\newlength{\wth}
\newcommand{\twographs}[2]
{\unitlength=1.1in
\begin{picture}(5.4,2.25) 
 \put(-0.75,-0.1){\epsfig{file=#1, width=1.05 \wth}}
\put(2.05,-0.1){\epsfig{file=#2, width=1.05 \wth}}
\put(-0.1,2.35){(a)}
\put(2.7,2.35){(b)}
\end{picture}
}

\newcommand{\twographsB}[2]
{\unitlength=1.1in
\begin{picture}(5.4,2)

 \put(0.0,0.1){\epsfig{file=#1, width=0.8 \wth, angle=0}}
\put(2.8,0.1){\epsfig{file=#2, width=0.8 \wth, angle=0}}
\put(0.0,2.15){(a)}
\put(2.8,2.15){(b)}
\end{picture}
}

\newcommand{\twographsC}[2]
{\unitlength=1.1in
\begin{picture}(5.4,2)

 \put(0.3,0.1){\epsfig{file=#1, width=0.7 \wth, angle=0}}
\put(3.1,0.1){\epsfig{file=#2, width=0.7 \wth, angle=0}}
\put(0.0,2.15){(a)}
\put(2.8,2.15){(b)}
\end{picture}
}

\newcommand{\AMSBthreegraphs}[3]{%
 \unitlength=0.5in
 \begin{picture}(12,4.2)
 \put(-1.5,0){\epsfig{file=#1, width=0.8\wth}}
 \put(2.7,0){\epsfig{file=#2, width=0.8\wth}}
 \put(6.85,0){\epsfig{file=#3, width=0.8\wth}}
 \put(-0.3,4.2){(a)}
 \put(4 ,4.2){(b)}
\put(8,4.2){(c)}
\end{picture}  
}

\newcommand{\sixgraphs}[6]{%
 \unitlength=1.1in
 \begin{picture}(5.4,5.2)(0,-0.2)
 \put(-0.0,4.35){\epsfig{file=#1, width=0.8\wth, angle=0}}
 \put(2.8,4.35){\epsfig{file=#2, width=0.8\wth,angle=0}}
 \put(0.0,6.2){(a)}
 \put(2.8,6.2){(b)}
 \put(-0.0,2.1){\epsfig{file=#3, width=0.8\wth,angle=0}}
 \put(2.8,2.1){\epsfig{file=#4, width=0.8\wth,angle=0}}
 \put(0.0,3.95){(c)}
 \put(2.8,3.95){(d)}
\put(0.1,-0.1){\epsfig{file=#5, width=0.8\wth,angle=0}}
\put(2.9,-0.1){\epsfig{file=#6, width=0.8\wth,angle=0}}
\put(0,1.9){(e)}
\put(2.8,1.9){(f)}
\end{picture}  
}

\newcommand{\sevengraphs}[7]{%
\unitlength=1.1in
\begin{picture}(5.2,5.9)(0,-0.5)
\put(-0.75,3.8){\epsfig{file=#1, width=1.0\wth}}
\put(1.85,3.8){\epsfig{file=#2, width=1.0\wth}}
\put(-0.15,6){(a)}
\put(2.45,6){(b)}
\put(-0.75,1.4){\epsfig{file=#3, width=1.0\wth}}
\put(1.85,1.4){\epsfig{file=#4, width=1.0\wth}}
\put(-0.15,3.6){(c)}
\put(2.45,3.6){(d)}
\put(-0.9,-0.5){\epsfig{file=#5, width=0.8\wth}}
\put(1.1,-0.5){\epsfig{file=#6, width=0.8\wth}}
\put(3.1,-0.5){\epsfig{file=#7, width=0.8\wth}}
\put(-0.4,1.3){(e)}
\put(1.65,1.3){(f)}
\put(3.65,1.3){(g)}
\end{picture}
}

\begin{document}

\begin{titlepage}

\hfill DAMTP-2009-37\\
 \vfil
\begin{center}
{\Large \bf Selecting a Model of Supersymmetry Breaking Mediation}
 \vspace{10mm}

S.S. AbdusSalam$^a$, B.C. Allanach$^a$, M.J. Dolan$^a$, F. Feroz$^b$ and M.P. Hobson$^b$
\\

\vspace{5mm}
\bigskip\medskip
\centerline{$^a$\it DAMTP, CMS, University of Cambridge, Wilberforce Road,}
\centerline{\it Cambridge CB3 0WA, United Kingdom}
\smallskip
\centerline{$^b$\it Cavendish Laboratory, University of Cambridge, JJ Thomson
  Avenue,}
\centerline{\it Cambridge, CB3 0HE, United Kingdom} 
\vfil
\end{center}
\begin{abstract}
\noindent
We study the problem of selecting between different mechanisms of
supersymmetry breaking in the Minimal Supersymmetric Standard Model using
current data. We evaluate the Bayesian evidence of four supersymmetry
breaking scenarios: mSUGRA, mGMSB, mAMSB and moduli mediation. The results
show a strong dependence on 
the dark matter assumption. Using the inferred cosmological relic density 
as an upper bound, minimal anomaly mediation is at least moderately favoured 
over the CMSSM\@. Our fits also
indicate that evidence for a positive sign of the $\mu$ 
parameter is moderate at best. 
We present constraints on the anomaly and gauge
mediated parameter spaces and some previously unexplored aspects of the dark
matter phenomenology of the moduli mediation scenario. 
We use sparticle searches, indirect observables and dark matter
observables in the global fit and quantify robustness with respect to prior
choice.  
We quantify how much information is contained within each constraint. 
\end{abstract}
\end{titlepage}

\tableofcontents

\section{Introduction}\label{sec:intro}
\paragraph{}
The Minimal Supersymmetric Standard Model (MSSM)\cite{Phys.Lett.B64.159,Nucl.Phys.B193.150} with R-parity can
solve the hierarchy problem and provide a candidate dark matter (DM)
particle. The absence of observed supersymmetric 
(SUSY) particles in collider experiments to date implies that if
nature is ever
supersymmetric then it must be broken. 
The source of SUSY breaking is a priori unknown in a bottom-up approach to
particle physics phenomenology. 
As spontaneous SUSY breaking in the MSSM is not viable~\cite{Nucl.Phys.B193.150}, one is led
to consider SUSY breaking in a 
hidden sector of the theory communicated via a messenger sector to the
MSSM. 
The choice of messenger sector leaves an imprint on the pattern of SUSY
breaking, and therefore on the expected SUSY phenomenology. There are
currently many viable soft-SUSY breaking schemes with some common mediation mechanisms including 
gauge, gravity, anomaly and moduli mediation.
There are examples of each class of model which only have a few parameters and
can be somewhat constrained by current cosmological and indirect collider data. 
Sparticles appear in loop contributions to electroweak and $B$-physics observables,
affecting their values indirectly. 
In this paper, we ask the question: is there sufficient power in such combined
data to favour one simple model over the other?
Such model
selection is likely to become even more interesting and important if signals
compatible with SUSY are found in the Large Hadron Collider (LHC) experiments. 

To date, global Bayesian fits have been performed to the constrained
MSSM (CMSSM)~\cite{Allanach:2005kz,Allanach:2006jc,hep-ph/0602028,hep-ph/0609295,arXiv:0705.0487,arXiv:0705.2012,arXiv:0806.1923,Trotta:2008bp,Feroz:2009dv,Trotta:2009gr}, 
large 
volume string compactification models (LVS)~\cite{Allanach:2008tu}, the
phenomenological MSSM (pMSSM)~\cite{AbdusSalam:2009qd} and the Non-Universal Higgs model~\cite{Roszkowski:2009sm}. In any Bayesian fit it is
essential to check for robustness by choosing widely different but reasonable
prior distributions of the model parameters. For models with more parameters
the prior dependence becomes greater, but more precise and direct data
will reduce such dependence. The pMSSM 
has {\em twenty}\/ extra-Standard Model (SM) parameters and displays a large
prior-dependence in the fit results\footnote{When we discuss the number of parameters additional to the SM
in a model, we refer explicitly to continuous parameters only.}. Even the CMSSM, with only
{\em four}\/ extra-SM parameters, shows significant prior dependence and 
so many aspects of the fits are not robust. 
The
parameter-space fits of the CMSSM and the pMSSM could only be robust with strong
enough 
direct and precise data. 
The LVS fit, which has
{\em two}\/ extra-SM parameters, shows approximate prior
independence~\cite{Allanach:2008tu}. 
Here, we shall perform fits to two
additional models: minimal anomaly mediated SUSY breaking (mAMSB) and minimal
gauge mediated SUSY breaking (mGMSB), each of which have the intermediate
number of {\em three}\/ parameters. Thus, we ask if the current
available data are powerful enough
to robustly constrain these models with three parameters additional to the SM
as well as ask if it is powerful enough to disfavour any of the models over
the others.
 The MSSM also
contains a $\mu$ parameter in the Higgs superpotential, whose magnitude is
constrained by the $Z$  boson mass but whose sign is unknown. 
In the past, MSSM Bayesian model selection has only focused on the extent to
which 
$\mu>0$ is favoured over
$\mu<0$~\cite{hep-ph/0609295,Allanach:2008tu,AbdusSalam:2009qd,Feroz:2008wr}. Thus,
our work extends 
the use of Bayesian model selection to cover different SUSY breaking
scenarios. 

Aside from the Bayesian fits, there have also been some global profile
likelihood (equivalent to minimizing 
$\chi^2$) analyses of the
CMSSM~\cite{arXiv:0705.0487,Trotta:2008bp,Buchmueller:2007zk,Buchmueller:2008qe}.  
Prior dependence does not appear in frequentist interpretations of fits 
and so one cannot be sure whether they are robust. The Bayesian analyses
indicate that such fits are not yet robust with current data. A
recent $\chi^2$ analysis~\cite{Heinemeyer:2008fb} compared the CMSSM, mGMSB and
mAMSB using electroweak 
and $B$-physics observables while omitting the dark matter constraint. The
analysis found that mAMSB was slightly preferred over the two other models by
$\Delta 
\chi^2=2$ and that light SUSY was slightly favoured by each model. We extend this
work 
in several ways, the most important being that 
we include dark matter as a constraint. We shall show that the dark matter
constraint contains more information content than the other observables.
By performing a Bayesian fit with several priors we are able to check
robustness of the fit, and, unlike Ref.~\cite{Heinemeyer:2008fb}, we incorporate
the uncertainties of important SM parameters into our fit.

The parameters of the CMSSM are a flavour
blind SUSY breaking scalar mass
$m_0$, a common gaugino mass $M_{1/2}$, a flavour blind SUSY breaking scalar trilinear coupling 
$A_0$ and $\tan \beta$, the ratio of the MSSM Higgs vacuum expectation
values (VEVs). Below a grand unification theory (GUT) scale of 
$ M_{GUT} \sim 2 \times 10^{16}$ GeV, the SUSY breaking terms of
different flavours evolve separately to the weak scale.
In anomaly mediated SUSY breaking (AMSB) \cite{Randall:1998uk}
SUSY-breaking is communicated to the
visible sector via the super-Weyl anomaly. 
In its original manifestation, pure anomaly mediation suffers from negative
slepton mass squared parameters, signalling a scalar potential minimum 
inconsistent with a massless photon. 
Minimal AMSB (mAMSB) assumes the existence of
an additional contribution to 
scalar masses $m_0$ at $M_{GUT}$
giving it a total of three 
parameters: the VEV of the auxiliary field in the supergravity multiplet
representing the overall sparticles mass scale, $m_{aux}$, $m_{0}$ and $\tan
\beta$.  
As advertised above, minimal gauge mediated SUSY breaking (mGMSB)~\cite{hep-ph/9801271}
also has three continuous parameters: the overall messenger mass scale,
$M_{mess}$, 
a visible sector soft SUSY-breaking mass scale, $\Lambda$ and $\tan\beta$.
It also contains an additional discrete parameter, namely $N_{mess}$, the
number of SU(5) $5\oplus \bar 5$ representations of mediating fields.  
The example of a moduli mediated model which we consider 
is the Large Volume Scenario (LVS) derived in the context of $IIB$ flux
compactification~\cite{hep-th/0502058, hep-th/0505076,hep-th/0610129, AbdusSalam:2007pm,Allanach:2008tu}, whose two extra-SM parameters
are an overall SUSY breaking mass scale $M_{3/2}$ and $\tan \beta$. 

The CMSSM is the phenomenologically most studied model in the literature, and so
it is useful as a benchmark as to how well other models fare in the fits. 
It is not clear without further model building how the GUT-scale flavour blind
structure of the CMSSM emerges, however. 
The main motivation for mAMSB and mGMSB is that the flavour problem is solved,
since gauge interactions in mGMSB necessarily lead to flavour blind soft
terms~\cite{hep-ph/9801271} 
at the messenger scale, and the mAMSB soft-terms are dominated by
flavour diagonal pieces proportional to the gauge
couplings~\cite{arXiv:0902.4880}. 
The main motivation for LVS is that it results from a string compactification
scenario which has moduli stabilization. 

The rest of the paper proceeds as follows: in Section~\ref{sec:analysis} we
show the technical 
aspects of our analysis, including a discussion of model parameters, prior probability distribution functions
(PDFs), observables and the calculation of the
likelihood. Section~\ref{sec:results:model_compar} presents  
an analysis of SUSY breaking model selection including  
the Bayesian evidence results on the preference for the sign of $\mu$, the
model selection and the dependence of our results on the dark matter
constraint. Sections~\ref{sec:fitgmsb} and~\ref{sec:fitamsb} detail the
phenomenologically viable regions of the 
parameter spaces of mGMSB and mAMSB, while Section~\ref{sec:con} discusses the
effects of 
individual observables and presents the best-fit points obtained, concluding
with a discussion of the interplay between the anomalous magnetic
moment of the muon and the rare branching ratio $BR(B\to X_s \gamma)$ in the
different models. Finally, we discuss the implications of our fits for the
direct detection of dark matter in Appendix~\ref{sec:DM}.

\subsection{Bayesian inference}\label{sec:bayesian}

Problems in data analysis generally divide into two categories: parameter estimation and model selection. In
parameter estimation problems one is interested in making inferences about the parameters of a given model using
the available data and any other prior information. Model selection problems are concerned with distinguishing
between different theories describing a given phenomenon. For instance, in the
case of the CMSSM, one would like to
know whether there is sufficient evidence in the data to rule out the $\mu<0$ branch. Bayesian inference provides
a consistent approach to model selection as well as to the estimation of a set
of parameters  $m$ in a
model (or hypothesis) $H$ for the data $d$. It can also be shown that Bayesian
inference is the unique 
consistent generalisation of the Boolean algebra~\cite{Am.J.Phys.vol.14.1-13}.

Assuming some model hypothesis $H$, 
Bayesian statistics helps update some
PDF $p(\underline m|H)$ of model parameters $\underline m$ with data. The
prior encodes our 
knowledge or 
prejudices about the parameters. Since $p(\underline m|H)$ is a PDF in
$\underline m$,
$\int p(\underline m|H) d\underline m=1$, which defines a normalization of the prior. One talks of
priors being `flat' in some parameters, but care must be taken to refer to the
measure of such parameters. A prior that is flat between some ranges in a
parameter $m_1$ will not be flat in a parameter $x \equiv \log m_1$, for
example.  
The impact of the data is encoded in the likelihood, or the PDF of obtaining
data set $\underline d$ from model point $\underline m$: $p(\underline
d|\underline m,H) \equiv {\mathcal L}(\underline m)$. 
The likelihood is a function of $\chi^2$, i.e.\ a statistical measure of how
well the data are fit by the model point. The desired quantity is the PDF of
the model parameters $m$ given some observed data $\underline d$ assuming hypothesis $H$:
$p(\underline m | \underline d, H)$.
Bayes' theorem states that
\begin{equation} p(\underline m|\underline d, H) =
\frac{p(\underline d|\underline m,H)p(\underline m|H)}
{p(\underline d|H)},
\label{eq:bayes}
\end{equation}
where $p(\underline d|H) \equiv \mathcal{Z}$ is the Bayesian evidence, the probability
density of observing data set $d$ integrated over all model parameter space.
The Bayesian evidence is given by: 
\begin{equation}
\mathcal{Z} =
\int{\mathcal{L}(\underline m)p(\underline m|H)}\ d \underline m
\label{eq:3}
\end{equation}
where the integral is over $N$ dimensions of the parameter
space $\underline m$. Since the Bayesian evidence is independent of the model
parameter values $\underline m$, it is usually ignored in parameter estimation
problems 
and posterior inferences are obtained by exploring the unnormalized
posterior using standard Markov Chain Monte Carlo sampling methods. 

In order to select between two models $H_{0}$ and $H_{1}$ one needs to compare
their respective posterior 
probabilities given the observed data set $\underline d$, as follows:
\begin{equation}
\frac{p(H_{1}|\underline d)}{p(H_{0}|\underline d)}
=\frac{p(\underline d|H_{1})p(H_{1})}{p(\underline d|
H_{0})p(H_{0})}
=\frac{\mathcal{Z}_1}{\mathcal{Z}_0}\frac{p(H_{1})}{p(H_{0})},
\label{eq:3.1}
\end{equation}
where $p(H_{1})/p(H_{0})$ is the prior probability ratio for the two models, which can often be set to unity
but occasionally requires further consideration (see e.g. \cite{arXiv:0811.1199,arXiv:0810.0781} for 
cases where the prior probability ratios should not be set to unity). It can be seen from Eq.~\ref{eq:3.1} that
Bayesian model selection revolves around the evaluation of the Bayesian evidence. As the average of likelihood
over the prior, the evidence automatically implements Occam's razor. 
A theory
with less parameters has a higher prior density since it integrates to
1 over the whole space.
Thus, there is an a priori preference for less parameters, unless the
data strongly require there be more.

Unfortunately, evaluation of Bayesian evidence involves the multi-dimensional
integral in Eq.~\ref{eq:3} and thus
presents a challenging numerical task. Standard techniques like thermodynamic integration \cite{O'Ruanaidh} are
extremely computationally expensive which makes evidence evaluation typically at least an order of magnitude more
costly than parameter estimation. Some fast approximate methods have been used for evidence evaluation, such as
treating the posterior as a multi-variate Gaussian centered at its peak (see e.g. \cite{astro-ph/0203259}), but this
approximation is clearly a poor one for highly non-Gaussian and multi--modal
posteriors, such as those we sample here.
Bridge sampling~\cite{radford,Bennett:2006fi,Stat.Sci.13163} allows the evaluation of the ratio of Bayesian evidence of
two models, but can yield rather imprecise results.
The problem can however be tackled by bank sampling \cite{arXiv:0705.0486},
but it is not yet clear how
precisely bank sampling can calculate the evidence ratio. 
Various alternative information criteria for model
selection are discussed by \cite{astro-ph/0701113}, but the evidence remains our
preferred method. 

The nested sampling approach, introduced in \cite{Skilling}, is a Monte Carlo method targeted at the 
efficient calculation of the evidence, but also produces posterior inferences as a by--product.
\cite{Feroz:2007kg,Feroz:2008xx} built on this nested sampling framework and introduced the {\sc MultiNest}
algorithm which is efficient in sampling from multi--modal posteriors that
exhibit curving degeneracies. {\sc MultiNest} produces
posterior samples and calculates the evidence and its uncertainty. This
technique has greatly reduced the 
computational cost of model selection and the exploration of highly degenerate multi--modal posterior
distributions. We employ this technique in this paper.

The natural logarithm of the ratio of posterior model probabilities provides a useful guide to what constitutes a
significant difference between two models:
\begin{equation}
\Delta \log \mathcal{Z} = \log \left[ \frac{p(H_{1}|\underline
    d)}{p(H_{0}|\underline d)}\right]
=\log \left[ \frac{\mathcal{Z}_1}{\mathcal{Z}_0}\frac{p(H_{1})}{p(H_{0})}\right].
\label{eq:Jeffreys}
\end{equation}
We summarize the convention we use in this paper in Table~\ref{tab:Jeffreys}.

\begin{table}
\begin{center}
\begin{tabular}{|c|c|c|c|}
\hline
$|\Delta \log \mathcal{Z}|$ & Odds & Probability & Remark \\ 
\hline\hline
$<1.0$ & $\lesssim 3:1$ & $<0.750$ & Inconclusive \\
$1.0$ & $\sim 3:1$ & $0.750$ & Weak Evidence \\
$2.5$ & $\sim 12:1$ & $0.923$ & Moderate Evidence \\
$5.0$ & $\sim 150:1$ & $0.993$ & Strong Evidence \\ \hline
\end{tabular}
\end{center}
\caption{The scale we use for the interpretation of model probabilities. Here the `$\log$'
represents the natural logarithm.}
\label{tab:Jeffreys}
\end{table}

\section{The Analysis} \label{sec:analysis}

\subsection{Choice of Prior Probability
Distributions}\label{sec:analysis:priors}

While for parameter estimation, the priors become irrelevant once the data are powerful enough, for model
selection the dependence on priors always remains (although with more informative data the degree of dependence
on the priors is expected to decrease, see e.g. \cite{0803.4089}); indeed this explicit dependence on priors
is one of the most attractive features of Bayesian model selection. Priors
should ideally represent one's state 
of knowledge before obtaining the data. Rather than seeking a unique `right'
prior, one should check the 
robustness of conclusions under reasonable variation of the priors. Such a
sensitivity analysis is required to 
ensure that the resulting model comparison is not overly dependent on a particular choice of prior and the
associated metric in parameter space, which controls the value of the integral involved in the computation of the
Bayesian evidence (for some relevant cautionary notes on the subject see \cite{Cousins:2008gf}). 

We have considered three different prior probability density functions (PDFs) in this analysis. The first is the standard ``linear prior'' where
\[p(m_1)=p(m_2) \]
for $m_{1,2}$ 
two different points in the parameter space of one of
the models under consideration. In particular, the linear priors are flat in
the ratio of the two MSSM Higgs vacuum expectation values (VEVs),
$\tan\beta$. It is important to realize that a prior which is flat in one set
of parameters $m$ is not necessarily flat in a different set of 
parameters $m'$, say. The two priors will be related by a Jacobian
factor 
such that 
\begin{equation}
 p(m)= \left| \frac{d \underline m'}{d \underline m} \right|
 p\mathbf({m'}). 
\end{equation}
One may consider a more fundamental set of parameters to be those involving
the quantities that actually appear in the Lagrangian before spontaneous
symmetry breaking. Such a set would be flat in the Higgs parameters $B$ and
$\mu$, but not in $\tan\beta$. ``Natural priors''~\cite{arXiv:0705.0487} are
priors which are flat in $B$ and $\mu$. The relationship between the two sets
of parameters is given by the conditions of electroweak symmetry breaking 
\begin{equation}
\mu B = \frac{\sin 2\beta}{2} \left( \bar{m}^2_{H_1} + \bar{m}^2_{H_2} +
2\mu^2\right) \ \textrm{~and~} \
\mu^2 = \frac{\bar{m}^2_{H_1}-\bar{m}^2_{H_2}\tan^2\beta}{\tan^2\beta -1} - \frac{M_Z^2}{2} ,
\end{equation}
which lead to the Jacobian factor~\cite{arXiv:0705.0487}
\begin{equation} 
J = \frac{M_Z}{2} \left| \frac{B}{\mu\tan\beta}
\frac{\tan^2\beta-1}{\tan^2\beta+1}\right|.
\label{eq:natprior}
\end{equation}
One may go further to examine the dependence upon the Higgs VEVs, Yukawa
couplings and renormalisation group effects in more sophisticated
priors~\cite{Cabrera:2008tj}. We instead focus on a few reasonable but
sufficiently 
different priors in order to check the prior independence of any inference
we make. 

It will also be the case in this analysis that we wish to estimate the scale
of a parameter, rather than its exact value. This is the case for the
messenger scale $M_{mess}$ in gauge mediation, for example. In this case the
appropriate prior to use is flat in $\log(m_1)$, and the relevant
Jacobian factor is proportional to $m_1^{-1}$. We call this the
``Jeffreys prior'' or simply the ``log prior''. 

\subsection{Parameters and ranges}\label{sec.param}

Before proceeding we specify the parameter ranges over which we
sample for the different models. Firstly, we consider both signs of $\mu$ in our analysis for  all
models. The ranges over which we vary the continuous model parameters are
shown in Table~\ref{tab:ranges1}. 
\begin{table}
\begin{center}
\begin{tabular}{|c|c|} \hline
CMSSM & mAMSB \\ \hline
 $50\mbox{~GeV}\leq m_0 \leq 4$ TeV & $50\mbox{~GeV}\leq m_0 \leq 4$ TeV \\
$50\mbox{~GeV} \leq m_{1/2} \leq 2$ TeV & $ 20\mbox{~TeV} \leq m_{3/2} \leq 200\mbox{~TeV}$  \\
$-4\mbox{~TeV} \leq A_0 \leq 4\mbox{~TeV}$ &   \\ \hline
\hline
mGMSB & LVS \\ \hline 
 $10^4\mbox{~GeV} \leq \Lambda \leq 10^5\mbox{~GeV}$  & $50\mbox{~GeV}\leq m_0\leq 2\mbox{~TeV}$ \\ 
$1.01\leq M_{mess}/\Lambda \leq 2\times 10^5$ & \\ \hline
\end{tabular}
\end{center}
\caption{Ranges for the parameters in mGMSB and the Large Volume Scenario. In
  mGMSB we also vary the discrete parameter $N_{mess}$ between 1 and 8.
For all models, $2\leq \tan \beta \leq 62$.}
\label{tab:ranges1}
\end{table}
$\tan \beta$ is bounded from below by 2, values lower than this are in
contravention of LEP2 Higgs searches, and from above by 62, since such large
values lead to non-perturbative Yukawa couplings below the GUT scale and
calculability is lost. 
In mGMSB the discrete parameter $N_{mess}$, the number of
messenger multiplets, is varied between 1 and 8. Higher values of
$N_{mess}$ lead to problems with perturbativity of gauge interactions at the
GUT scale \cite{hep-ph/9801271}. In the CMSSM the unification scale is the standard GUT scale $m_{GUT} \approx 2\times 10^{16}$GeV, while for 
the LVS the soft terms are defined at the intermediate string  scale $m_s \approx 10^{11}$GeV as in~\cite{Allanach:2008tu}.

\subsection{The likelihood}\label{sec:analysis:like}

\begin{table} 
\begin{center}
\begin{tabular}{|c|c|c|c|} \hline
Observable & Constraint & Theory & Experiment  \\ \hline \hline
$m_W$[GeV]              & $80.399\pm0.027$      & \cite{hep-ph/0604147,arXiv:0710.2972} & \cite{arXiv:0808.0147}   \\ \hline
$\sin^2 \theta^l_{eff}$ & $0.23149\pm0.000173$  & \cite{hep-ph/0604147,arXiv:0710.2972} & \cite{arXiv:0712.0929}   \\ \hline
$\delta a_{\mu}\times 10^{10}$ & $29.5\pm8.8$   & \cite{Belanger:2008sj,Belanger:2006is,Belanger:2004yn,Belanger:2001fz,hep-ph/0405255, hep-ph/0312264,hep-ph/0609168,arXiv:0808.1530} & \cite{Phys.Lett.B667.1}   \\ \hline
$\Omega_{\rm DM}h^2$  &  $0.1143\pm 0.02$ & \cite{Belanger:2008sj,Belanger:2006is,Belanger:2004yn,Belanger:2001fz} & \cite{arXiv:0803.0547} \\ \hline
$m_h$[GeV]              & $>114.4$GeV & \cite{hep-ph/0406166} & \cite{hep-ex/0306033} \\ \hline \hline
$\Gamma_Z^{tot}$[GeV]      & $2.4952\pm0.0023$ & \cite{hep-ph/0604147} & \cite{hep-ex/0509008, yellowbook} \\ \hline
$R^0_l$                &  $20.767\pm0.025$ & \cite{hep-ph/0604147} & \cite{hep-ex/0509008, yellowbook} \\ \hline
$R^0_b$                 & $0.21629\pm0.00066$ & \cite{hep-ph/0604147} & \cite{hep-ex/0509008, yellowbook} \\ \hline
$R^0_c$                 &$0.1721\pm0.0030$ & \cite{hep-ph/0604147} & \cite{hep-ex/0509008, yellowbook} \\ \hline
$A^{0,b}_{fb}$         &$0.0992\pm0.0016$ & \cite{hep-ph/0604147} & \cite{hep-ex/0509008, yellowbook} \\ \hline
$A^{0,c}_{fb}$         &$0.0707\pm0.035$ & \cite{hep-ph/0604147} & \cite{hep-ex/0509008, yellowbook} \\ \hline
$A_{LR}^0(SLD)$             & $ 0.1513\pm0.0021$ & \cite{hep-ph/0604147} & \cite{hep-ex/0509008, yellowbook} \\ \hline
$\mathcal{A}_b$        &$0.923\pm0.020$ & \cite{hep-ph/0604147} & \cite{hep-ex/0509008, yellowbook} \\ \hline
$\mathcal{A}_c$         &$0.670\pm0.027$ & \cite{hep-ph/0604147} & \cite{hep-ex/0509008, yellowbook} \\ \hline \hline

$BR(B\to X_s \gamma)\times 10^4$ & $ 3.52\pm 0.39 $ & \cite{arXiv:0710.2067, arXiv:0808.3144} & \cite{arXiv:0808.1297}   \\ \hline
$BR(B_s\to\mu^+\mu^-)$ & $<5.8\times10^{-8}$ & \cite{Belanger:2008sj,Belanger:2006is,Belanger:2004yn,Belanger:2001fz,lin} & \cite{arXiv:0712.1708} \\ \hline
$BR(B\to D \tau \nu)$   &$0.416\pm0.138$ & \cite{arXiv:0710.2067, arXiv:0808.3144} & \cite{arXiv:0709.1698} \\ \hline 
$R_{\Delta M_s}$ & $0.85\pm0.11$ & \cite{hep-ph/0605012,hep-ph/0210145} & \cite{hep-ph-0606167,arXiv:0808.1297} \\ \hline
$R_{B\tau\nu}$      & $1.28\pm0.40$ & \cite{arXiv:0710.2067, arXiv:0808.3144,hep-ph/0605012} &  \cite{arXiv:0808.1297} \\ \hline
$\Delta_{0-}$      & $0.031^{+0.03}_{-0.025}$ & \cite{arXiv:0710.2067, arXiv:0808.3144} & \cite{Phys.Lett.B667.1,arXiv:0808.1915,hep-ex/0402042} \\ \hline
$R_{l23}$               &$1.004\pm0.007$ & \cite{arXiv:0710.2067, arXiv:0808.3144}
      & \cite{arXiv:0801.1817} \\ \hline \hline 
$m_t$[GeV]              & $172.4\pm1.2$         & - & \cite{arXiv:0808.1089}   \\ \hline
$m_b(m_b)^{\overline{MS}}$[GeV]              & $4.20\pm0.07$         & - & \cite{Phys.Lett.B667.1} \\ \hline
$m_Z$[GeV]              & $91.1876\pm0.0023$    & - & \cite{Phys.Lett.B667.1,hep-ex/0509008, yellowbook}   \\ \hline
$\alpha_s^{\overline{MS}}(M_Z)$         &$0.1172\pm0.002$       & - & \cite{arXiv:0803.0342} \\ \hline
$1/\alpha^{\overline{MS}}$     & $127.918\pm0.018$     & - & \cite{Phys.Lett.B667.1} \\ \hline 
\end{tabular}
\end{center}
\caption{Experimental constraints, showing the observables, the constraints applied and the source of the theoretical and experimental values and errors. The first section shows general observables, the second electroweak observables, the third
section B-physics constraints and the final section contains SM parameter
constraints. The 95$\%$ confidence level (CL) direct sparticle search constraints
from \cite{Phys.Lett.B667.1} were also applied where relevant.}
\label{tab:obs}
\end{table}

The calculation of the likelihood follows \cite{arXiv:0705.0487} with updated data and additional observables. The
constraints we use are all shown in Table~\ref{tab:obs} along with their respective experimental and theoretical
sources. We treat the measurements $D_i$ of the observables as independent and we have been careful to eliminate
possible correlations between them. We also assume Gaussian errors on all
measurements except where explicitly mentioned. The log likelihood 
of a prediction $p_i$ of an observable $i$ is given by
\begin{equation}
\log \mathcal{L}_i = -\frac{\chi^2_i}{2} -\frac{1}{2}\log(2\pi) - \log(\sigma_i)
\end{equation}
where $\chi_i^2 = \frac{(c_i-p_i)^2}{\sigma_i^2} $, $c_i$ is the central
experimental value and $\sigma_i$ is 
the standard deviation incorporating both experimental and theoretical
uncertainties. All likelihoods except for 
those of the DM relic density, the Higgs mass and
$BR(B_s\to\mu^+\mu^-)$ are calculated in this way. Details of the
constraints applied on the last two exceptions mentioned above
can be found in
\cite{Feroz:2008wr}. The ${\mathcal L}_i$ considered here are independent, and
are multiplied together to calculate the overall likelihood. 

\begin{figure}
\begin{center}
\begin{picture}(0,0)%
\includegraphics{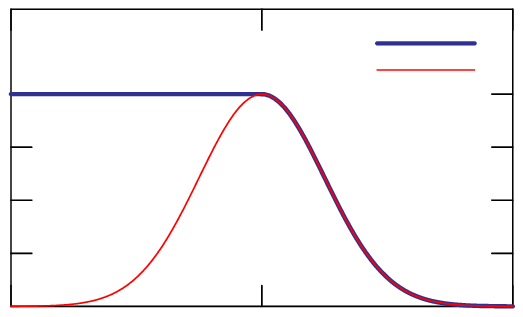}%
\end{picture}%
\setlength{\unitlength}{0.0200bp}%
\begin{picture}(10800,6480)(0,0)%
\put(2475,2414){\makebox(0,0)[r]{\strut{} 0.25}}%
\put(2475,3179){\makebox(0,0)[r]{\strut{} 0.5}}%
\put(2475,3943){\makebox(0,0)[r]{\strut{} 0.75}}%
\put(2475,4707){\makebox(0,0)[r]{\strut{} 1}}%
\put(2750,1100){\makebox(0,0){\strut{} 0.0343}}%
\put(6363,1100){\makebox(0,0){\strut{} 0.1143}}%
\put(9975,1100){\makebox(0,0){\strut{} 0.1943}}%
\put(550,3790){\rotatebox{90}{\makebox(0,0){\strut{}$L/L_{max}$}}}%
\put(6362,275){\makebox(0,0){\strut{}$\Omega_{DM} h^2$}}%
\put(7750,5438){\makebox(0,0)[r]{\strut{}constraint}}%
\put(7750,5054){\makebox(0,0)[r]{\strut{}pure WMAP5}}%
\end{picture}%
\caption{Depiction of our likelihood constraint on the predicted value of
  $\Omega_{\rm DM}h^2$ due to the lightest
neutralino, compared with a simple Gaussian with WMAP5 central value and a 1$\sigma$ uncertainty of 0.02.}
\label{fig:omega}
\end{center}
\end{figure}

For $\Omega_{\rm DM}h^2$, we use two different forms of constraint. Assuming
that the DM relic density is 
composed entirely of the lightest SUSY particle (LSP), one obtains a Gaussian $\mathcal{L}_{\rm DM}$
with the WMAP5 central value. We call this the `symmetric $\mathcal{L}_{\rm DM}$'. 
It is of course possible that the neutralino does not make up the entirety of
the DM in the universe. For instance, axions or stable moduli could
make up the additional component.
Indeed, in the Large Volume Scenario, the properties of the lightest K\"ahler
modulus as a DM constituent have been discussed in~\cite{arXiv:0705.3460}.  In mAMSB, the wino co-annihilation is so efficient that the relic density is generically far
below the WMAP value, so that such an extra component is cosmologically necessary. This motivates the use of the
`asymmetric $\mathcal{L}_{\rm DM}$' which is given as:
\begin{equation}
\mathcal{L}_{\rm DM}(x \equiv \Omega_{\rm DM}h^2) = 
\begin{cases}
\frac{1}{c+\sqrt{\pi s^2/2}}, & \mbox{if $x < c$} \\
\frac{1}{c+\sqrt{\pi s^2/2}}\exp\left[-\frac{(x-c)^2}{2s^2}\right], & \mbox{if $x \geq c$}. \\
\end{cases}
\label{omega_like}
\end{equation}
where $c$ and $s$ are the mean and 1-$\sigma$ error values of the Gaussian
$\mathcal{L}_{\rm DM}$. We take 
$c=0.1143$ that results from a fit to WMAP5, baryon acoustic oscillations and Type Ia
supernovae 
observations. We also take $s=0.02$ in order to incorporate an estimate of 
higher order uncertainties in its prediction~\cite{Baro:2007em}. For the asymmetric $\mathcal{L}_{\rm DM}$ we assume that the other dark matter component is R-parity even, so
therefore we apply the constraint that the LSP must not be charged or
coloured, since then it would be stable and have appeared in rare isotope
searches.
A diagram of
the resulting likelihood penalty is displayed in Fig.~\ref{fig:omega}.

In mGMSB the gravitino is the LSP\@. The gravitino mass
$m_{3/2}$ depends on the scale of SUSY breaking and is usually in the range of a few eV up to a few
hundred keV. This property has often been cited as an attractive feature of mGMSB theories, as the low mass of the
gravitino indicates that gravity mediated effects are much smaller than gauge mediated effects, so that flavour
changing neutral currents are naturally suppressed. The gravitino contribution to the relic density is given by~\cite{Phys.Rev.Lett.48.223}
\begin{equation}
 \Omega_{3/2}h^2 \approx \frac{m_{3/2}}{\mbox{keV}}\left[\frac{100}{g_{*}(T_f)} \right]
\label{eq:m32density}
\end{equation}
where $g_*(T_f)$ is the number of massless degrees of freedom at the gravitino freeze-out temperature, and for
SUSY models is in the range 100-200~\cite{hep-ph/9801271}. We see from Eq.~\ref{eq:m32density} that achieving the
WMAP value of the relic density requires a gravitino at the lower end of the mass range, in particular less
than 1 keV. Constraints on structure formation and
WMAP~\cite{astro-ph/0303622} are  now strong enough, however, to rule
out the resultant warm DM\@. Heavier gravitinos which evade this bound
would be over abundant compared to observation. Although it may be possible to
dilute the gravitino relic density by late entropy production,
we wish to keep our analysis as general 
as possible without confining ourselves to a specific change in the physics of the early universe. Accordingly,
in our analysis we do not impose any DM constraint on mGMSB.

The combined log likelihood is the sum of the individual likelihoods,
\begin{equation}
\log\mathcal{L}^{tot} = \sum_i \log\mathcal{L}_i. 
\end{equation}
To calculate the MSSM spectrum we use \texttt{Softsusy2.0.18}~\cite{Allanach:2001kg}
which calculates the spectrum of the CMSSM, mAMSB and mGMSB\@. By modifying the
unification scale from $m_{GUT}$ to $m_{string}\sim 10^{11}$GeV and the gauge
coupling boundary conditions \texttt{Softsusy} can also provide the spectrum
in the LVS case. Parameter space points which violate the current sparticle
exclusion bounds of~\cite{Phys.Lett.B667.1}, do not break electroweak 
symmetry correctly or have tachyonic sparticles are assigned zero likelihood. Points which have a charged LSP are
rejected\footnote{Due to the small neutralino-chargino splitting in mAMSB we
  have been careful 
to reject any points that would violate the long-lived charged stable particle
bounds from Tevatron, which requires $\Delta m= m_{\chi_1^+}-m_{\chi_1^0} >
50$ MeV. In fact, we find that this bound does not constrain the mAMSB
parameter space
since mAMSB predicts larger splittings~\protect\cite{hep-ph/9904378}.}. 
If a point survives the cuts above, it is passed via the SUSY Les
Houches Accord~\cite{hep-ph/0311123} to \texttt{microMEGAS2.2}~\cite{Belanger:2008sj,Belanger:2006is,Belanger:2004yn,Belanger:2001fz}, \texttt{SuperIso2.3}~\cite{arXiv:0710.2067, arXiv:0808.3144}
and \texttt{SusyPOPE}~\cite{arXiv:0710.2972}. From \texttt{microMEGAS} we obtain the
DM relic density, the rare
branching ratio $BR(B_s\to\mu^+\mu^-)$, the SUSY component $\delta a_\mu$ of
the anomalous magnetic moment of the 
muon $(g-2)_{\mu}$ and DM direct detection rates. To the one-loop value of
$a_{\mu}$ calculated by {\tt microMEGAS}, we add the logarithmic 
piece of the quantum electro-dynamics 2-loop calculation, the 2-loop stop-Higgs and chargino-stop/bottom pieces and the recently
discovered two-loop effect due to a shift in the muon Yukawa coupling and
proportional to $\tan^2\beta$~\cite{hep-ph/0405255, hep-ph/0312264,hep-ph/0609168,arXiv:0808.1530}. From
\texttt{SuperIso2.3} the branching ratios $BR(B\to X_s\gamma)$, $BR(B\to 
D\tau\nu)$, the quantities $R_{\Delta M_s}$, $R_{l23}$, $R_{B\tau\nu}$ and the
isospin asymmetry $\Delta_{0-}$ 
are obtained\footnote{We note that in the process of preparing this paper and
  after our fits were performed a new version of
  \texttt{SusyBSG}\cite{arXiv:0712.3265} appeared. This more accurate
  calculation could result in a change in our $BR(B\to X_s \gamma)$ prediction
  of up to
  $0.13\times 10^{-4}$.}. \texttt{SusyPOPE} provides all of the observables
listed in the second part of Table~\ref{tab:obs}.

We present our data in the form of
marginal posteriors in one and two dimensions, where the unseen dimensions
have been integrated over. To do this integration 
numerically we divide the range of the parameter in which we are interested
into a series of bins, and then count 
the number of samples which fall into each bin.  We use 60 bins per parameter,
which is a trade-off between parameter resolution and unwanted
statistical noise from finite sampling effects.
We shall also discuss the profile likelihood at various stages. The profile
likelihood is often shown in one dimension, and plots the maximum of the
likelihoods of the samples that fall in each bin. This statistic is equivalent
to plotting minimum $\chi^2$ in various directions of parameter space, 
and shows where the data are fitted best, but does not take into account
volume effects of the less best-fit points. Everywhere in this paper, we take
best-fit to mean the highest likelihood (or equivalently, lowest $\chi^2$). 
{\sc MultiNest} is optimized for
calculating Bayesian evidence values, and also typically gives a reasonable
sampling for posterior evaluation. However, the profile likelihood is
extracted by {\sc MultiNest} with a lot of noise. It was found by
Ref.~\cite{Trotta:2008bp} to depend upon the prior, 
which it should not do for
a large enough sampling. 
Therefore, here we do not explicitly
include plots of the profile likelihood, preferring instead to mention its
behaviour at various points in the text.  

We have taken care that all the two-sided constraints applied in the likelihood are normalised to 1. For one-sided constraints such as sparticle exclusion limits,
the likelihood is zero below the 95\% confidence limit, and 1 above that limit. The likelihoods we apply for $BR(B_s\to\mu^+\mu^-)$ and $m_h$ 
are more complicated, but they both asymptote to  1 in the limit of zero branching ratio and large Higgs mass respectively. While we focus on the 
statistically meaningful {\em difference} between evidences in our fits, by specifying the normalisation of the likelihood and providing 
also the absolute values of the evidence in our tables of results  we hope to encourage other groups, by following our calculation, to directly compare Bayesian evidences
 calculated for other models with the results presented herein.

\label{sec:results}

\section{Model Selection}\label{sec:results:model_compar}

In this section we discuss whether current data prefer one model over the
others in the set  mAMSB,
mGMSB, the CMSSM and LVS. 
We also present our results on quantification of the preference of the fits
for $\mu > 0$. In both these cases, we do the analysis with `symmetric $\mathcal{L}_{\rm DM}$' as well
as `asymmetric $\mathcal{L}_{\rm DM}$' (see
Section~\ref{sec:analysis:like}). 
Strictly speaking, a calculation of the Bayesian evidence with asymmetric dark
matter constraints is inherently unfair unless the relic density of the 
additional DM component is included, with an associated reduction in
the evidence due to the Occam's razor effect. We shall here make model
selections between models where the asymmetric likelihood penalty has been
applied to all of the models in question. Thus, although the Occam's razor
effect is neglected, it should affect all of the models being compared in
approximately the same way and so the comparison should remain unaffected. 
We also assume that the prior
probabilities of each model are equal and independent of the sign of $\mu$.

\subsection{A preference for mAMSB, mGMSB, the CMSSM or LVS?}\label{sec:results:model_compar:models}

\begin{table*}
\begin{center}
\begin{tabular}{|c|c|c|c|c|c|c|}
\hline
& \multicolumn{3}{c|}{symmetric $\mathcal{L}_{\rm DM}$} & \multicolumn{3}{c|}{asymmetric $\mathcal{L}_{\rm DM}$} \\
\hline\hline
Model/Prior  & linear & log & natural & linear & log & natural \\
\hline
CMSSM & $ 8.0 \pm 0.1 $ & $ 7.9 \pm 0.1 $ & $ 10.3 \pm 0.1 $ & $ 0.0 \pm 0.1 $ & $ 1.0 \pm 0.1 $ & $ 1.3 \pm 0.1 $ \\ 
mAMSB  & $ 0.4 \pm 0.1 $ & $ 0.6 \pm 0.1 $ & $ 0.0 \pm 0.1 $ & $ 5.1 \pm 0.1 $ & $ 6.0 \pm 0.1 $ & $ 5.0 \pm 0.1 $ \\ 
LVS    & $ 8.7 \pm 0.1 $ & $ 8.9 \pm 0.1 $ & $ 11.8 \pm 0.1 $ & $ 2.9 \pm 0.1 $ & $ 3.0 \pm 0.1 $ & $ 3.1 \pm 0.1 $ \\ 
\hline
\end{tabular}
\caption{log evidences ($\Delta \log \mathcal{Z}$) for mAMSB, LVS and the CMSSM
  for both signs of $\mu$. 
Symmetric $\mathcal{L}_{\rm DM}$ labels the assumption that the DM relic density is composed entirely
of the LSP and asymmetric $\mathcal{L}_{\rm DM}$ labels the assumption that
the LSP forms only a part of the 
DM relic density. The log evidence of the natural prior mAMSB, $\log Z_s = 67.3$ and the log evidence of the linear prior CMSSM, 76.7 have been subtracted 
from all entries in the symmetric $\mathcal{L}_{\rm DM}$ and asymmetric $\mathcal{L}_{\rm DM}$ respectively.}
\label{tab:model-prob-odds}
\end{center}
\end{table*}

We now discuss to what extent the mAMSB, mGMSB, the CMSSM, and LVS model are
preferred over one another 
by the current data. In this subsection, we marginalise over the sign of $\mu$. 

The log evidence values ($\log
\mathcal{Z}$) for mAMSB, LVS and the CMSSM are the most important results
of this paper and they are shown
in Table~\ref{tab:model-prob-odds}.
mAMSB is strongly  favoured over the
CMSSM ($\Delta \log \mathcal{Z} > 5$) and moderately favoured over LVS ($\log
\Delta \mathcal{Z} > 2$) for asymmetric $\mathcal{L}_{\rm DM}$, a result which
is approximately prior independent. 
However, mAMSB is almost decisively ruled
out for symmetric 
$\mathcal{L}_{\rm DM}$ ($\Delta \log \mathcal{Z} < -7$). 
Although mAMSB
with a purely thermal relic density is decisively disfavoured we have not taken
into account the non-thermal component of the relic density due to decays of
the heavy gravitino in mAMSB which for, some values of the inflationary
reheating temperature, could saturate the WMAP bounds. 
As mentioned above,
in mAMSB, the degeneracy between the LSP and next-to-LSP (NLSP) chargino
causes extremely 
efficient co-annihilation of the LSP to near zero relic density over the whole
parameter space. 
Thus, the entire parameter space of mAMSB passes the asymmetric ${\mathcal
  L}_{\rm DM}$ constraint and it is strongly preferred. On
the other hand the LSP relic density in most of the mAMSB parameter space is
much lower than the central value inferred by 
WMAP and hence mAMSB is almost ruled out if one assumes that the DM relic
density is composed entirely of the LSP.
There is a weak to moderate
preference for LVS over the CMSSM depending on the prior and whether the
symmetric or 
asymmetric form of  $\mathcal{L}_{\rm DM}$ is taken.

\begin{table*}
\begin{center}
\begin{tabular}{|c|c|c|c|}
\hline
Model/Prior  & linear & log & natural \\
\hline\hline
CMSSM  & $ 0.9 \pm 0.1 $ & $ 1.0 \pm 0.1 $ & $ 1.1 \pm 0.1 $ \\ 
mAMSB  & $ 1.1 \pm 0.1 $ & $ 1.2 \pm 0.1 $ & $ 2.9 \pm 0.1 $ \\ 
mGMSB & $ 1.4 \pm 0.1 $ & $ 1.5 \pm 0.1 $ & $ 0.0 \pm 0.1 $ \\
LVS    & $ 1.1 \pm 0.1 $ & $ 0.8 \pm 0.1 $ & $ 1.5 \pm 0.1 $ \\ 
\hline
\end{tabular}
\caption{log evidences ($\log \mathcal{Z}$) for mAMSB, mGMSB, LVS and the CMSSM for both signs of $\mu$ and
without imposing the $\Omega_{\rm DM}h^2$ constraint. A constant value (the
log evidence of mGMSB using  
natural priors, 78.2) has been subtracted from all the log evidence values.}
\label{tab:model-noDM-prob-odds}
\end{center}
\end{table*}

To quantify the extent to which these results depend on the DM constraint, we calculate the
Bayesian evidence ratios for mAMSB, mGMSB, LVS and the CMSSM 
without this constraint and list them in
Table~\ref{tab:model-noDM-prob-odds}. It is clear from these evidence values that the results are completely
inconclusive in the absence of the DM constraint and hence it can be
concluded that it does indeed dominate our model selection results. 
Ref.~\cite{Heinemeyer:2008fb} concluded on the basis of a $\chi^2$
minimisation with a subset of the observables included in the present paper
and no DM constraint
that mAMSB was  slightly preferred over the other two models. 
Table~\ref{tab:model-noDM-prob-odds} indicates that even such a weak inference
cannot yet be made due to non-robustness of fits in mAMSB, since the
preference for or against it depends upon the form of the prior.

\subsection{Selection of sign$(\mu)$}\label{sec:results:model_compar:sign_mu}

\begin{table*}
\begin{center}
\begin{tabular}{|c|c|c|c|c|c|c|}
\hline
& \multicolumn{3}{c|}{symmetric $\mathcal{L}_{\rm DM}$} & \multicolumn{3}{c|}{asymmetric $\mathcal{L}_{\rm DM}$} \\
\hline\hline
Model/Prior  & linear & log & natural & linear & log & natural \\
\hline
CMSSM & $ 1.2 \pm 0.1 $ & $ 2.4 \pm 0.1 $ & $ 0.4 \pm 0.1 $ & $ 1.3 \pm 0.1 $ & $ 2.4 \pm 0.1 $ & $ 0.4 \pm 0.1 $ \\ 
mAMSB  & $ 1.4 \pm 0.1 $ & $ 2.5 \pm 0.1 $ & $ 0.4 \pm 0.1 $ & $ 1.9 \pm 0.1 $ & $ 3.4 \pm 0.1 $ & $ 0.6 \pm 0.1 $ \\ 
LVS    & $ 3.2 \pm 0.1 $ & $ 3.1 \pm 0.1 $ & $ 2.6 \pm 0.1 $ & $ 3.6 \pm 0.1 $ & $ 3.9 \pm 0.1 $ & $ 3.3 \pm 0.1 $ \\ 
\hline\hline
& \multicolumn{3}{c|}{} & \multicolumn{3}{c|}{No DM constraint} \\ \hline mGMSB & \multicolumn{3}{c|}{} & $1.7\pm0.1$& $2.2\pm0.1$& $1.4\pm0.1$\\
\hline
\end{tabular}
\caption{Difference in log evidences ($\Delta \log \mathcal{Z}$) for $\mu > 0$
  and $\mu < 0$ for the CMSSM, mAMSB and the
LVS model. A positive value for a model indicates a preference for $\mu > 0$. `Symmetric $\mathcal{L}_{\rm DM}$' labels the assumption that the DM
relic density is composed entirely of the LSP and `asymmetric $\mathcal{L}_{\rm DM}$' labels the assumption that
the LSP forms only a part of the DM relic density.}
\label{tab:sign_prob_odds}
\end{center}
\end{table*}

We summarize the amount of preference for $\mu > 0$ in mAMSB, the CMSSM and LVS
in 
Table~\ref{tab:sign_prob_odds}. We list the Bayes factors, $\Delta \log \mathcal{Z}$ in favour of $\mu >
0$ for these three models. Using the Jeffreys scale defined in Table~\ref{tab:Jeffreys},
it is evident that although there is a positive
but not strong evidence in favour of $\mu > 0$ for all three models, the extent
of the preference depends 
quite strongly on the priors used.  
Although there are small numerical
differences due to the constraints being updated and different prior ranges
taken, the CMSSM $\Delta \log 
{\mathcal Z}$ values lead to the same conclusions as previous
determinations~\cite{hep-ph/0609295,Allanach:2008tu,AbdusSalam:2009qd,Feroz:2008wr}:
the preference for $\mu>0$ is moderate at best, and quite prior dependent. 
This dependence on the prior is a clear sign that the data are not yet of
sufficiently high quality to be able to distinguish between these models
unambiguously and hence it is not justified to ignore the $\mu < 0$ branch in
any of the models under consideration, despite the fact that $(g-2)_\mu$
favours positive $\mu$ by around 3.4$\sigma$. 

\section{mGMSB Parameter Constraints \label{sec:fitgmsb}}

\begin{figure*}
\begin{center}
\twographsB{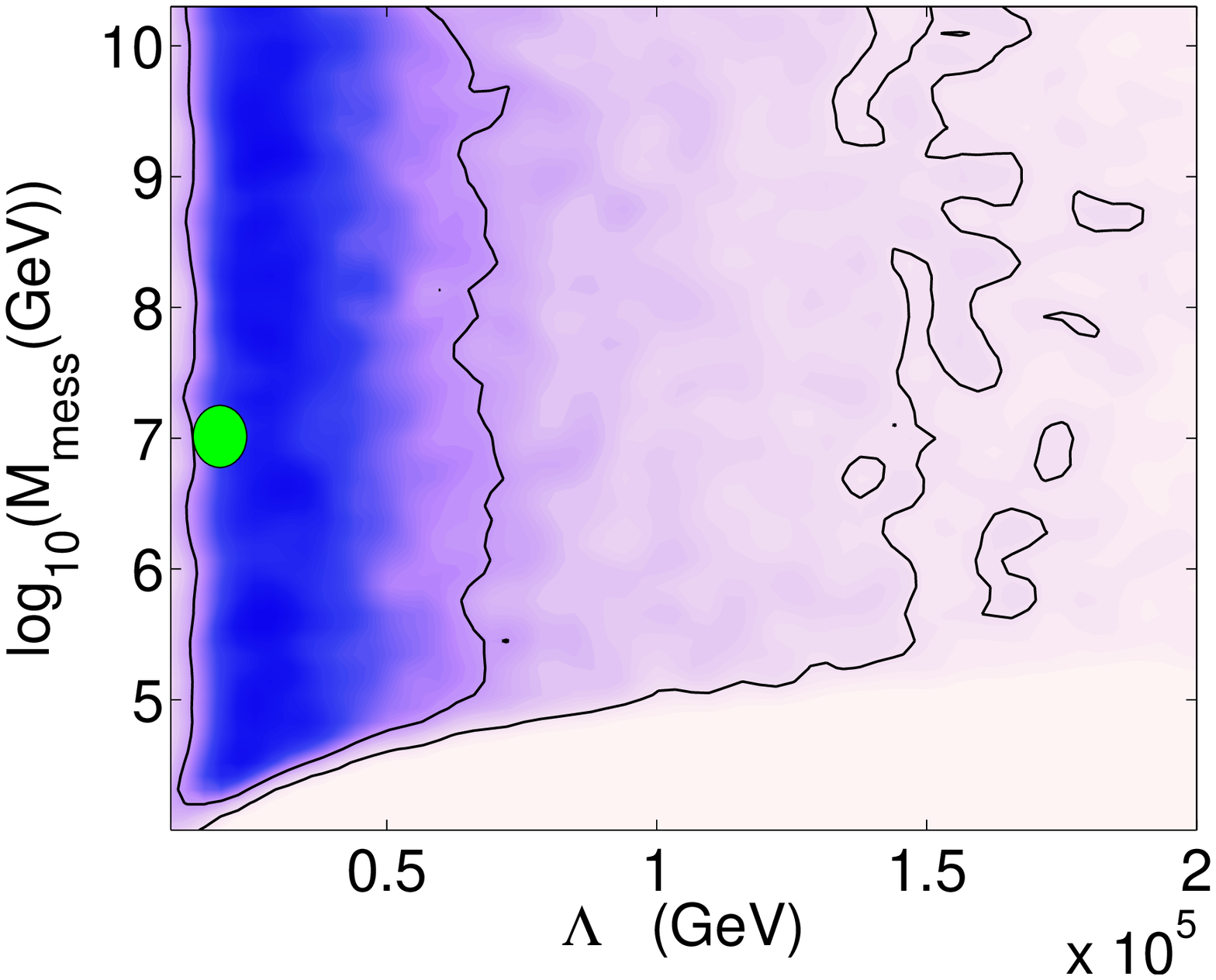}{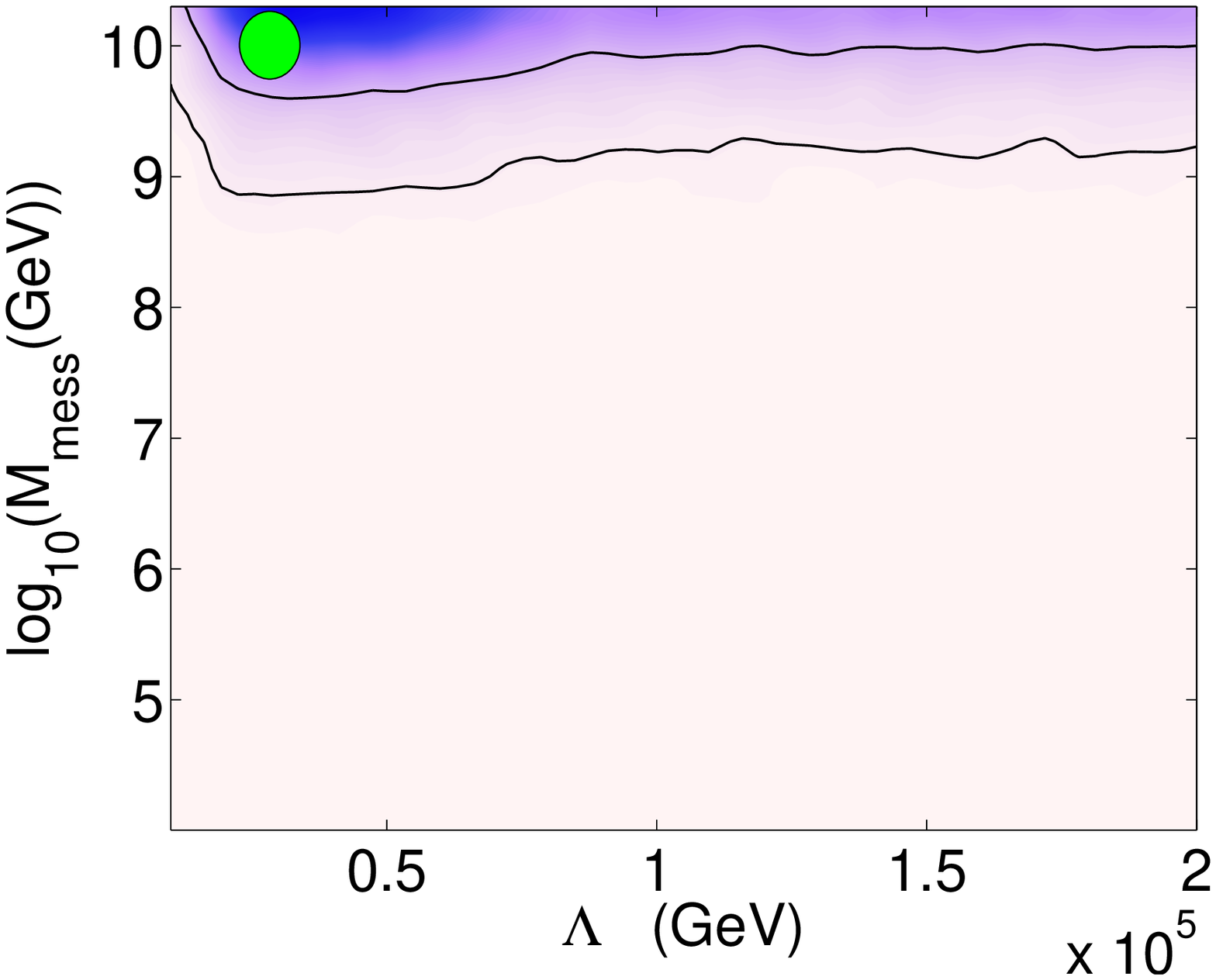}
\caption{2D marginalised posterior PDFs for mGMSB in the
  $\log_{10}(M_{mess})$-$\Lambda$ plane for (a) log priors and (b) natural
  priors with 68\% and 95\% Bayesian credibility intervals. Both plots are marginalized over both signs of $\mu$.  Green dots show the
  position of the best-fit point for each prior choice sample.} 
\label{fig:mGMSB2d}
\end{center}
\end{figure*}

\begin{figure}
\begin{center}
\twographs{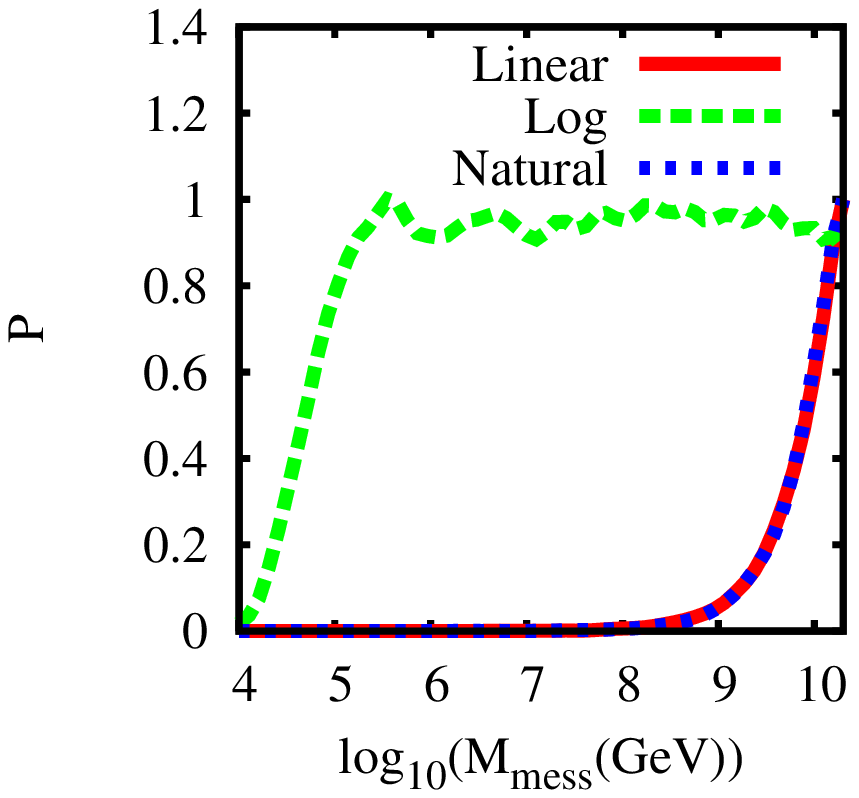}{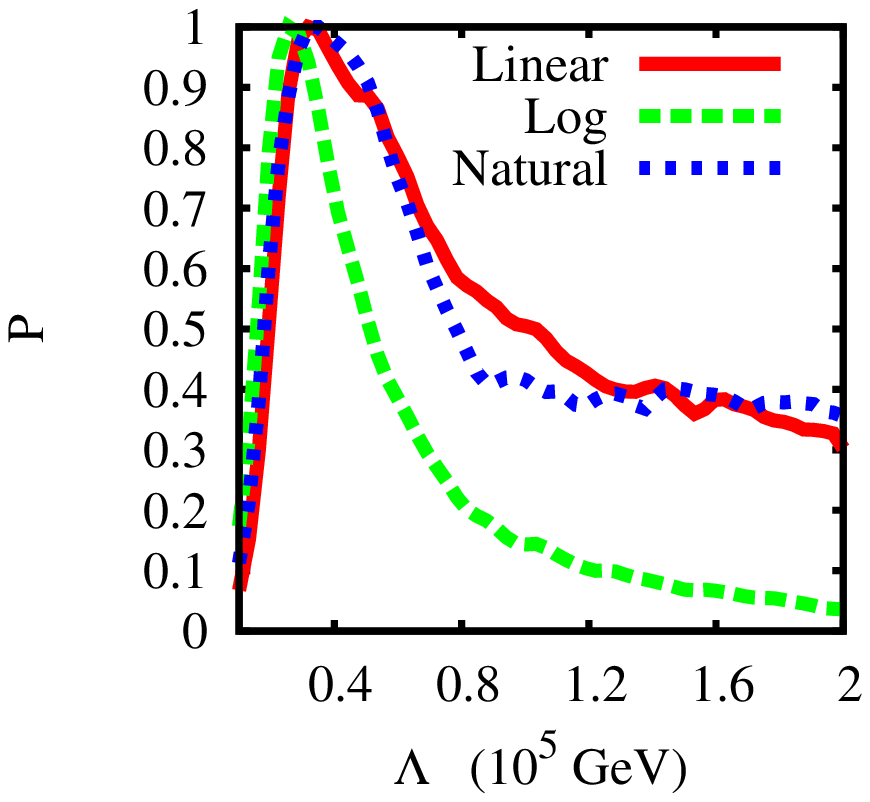}
\caption{mGMSB 1D marginalized posterior PDFs over
(a)  $\log_{10}(M_{mess})$ and (b) $\Lambda$ 
for linear, log
  and natural priors. Both plots are marginalized over both signs of $\mu$. 
Both here and in future one-dimensional marginalizations, the vertical axis
has been normalized so that the maximum posterior PDF is one.
}
\label{fig:mGMSB1d}
\end{center}
\end{figure}
We now discuss the constraints on mGMSB coming from the global fit.
These have not appeared in the literature before, except for in
Ref.~\cite{Heinemeyer:2008fb}, where a $\chi^2$ minimisation was performed,
and so prior dependence was not quantified. 
Fig.~\ref{fig:mGMSB2d} shows the marginalized joint posterior PDFs for
mGMSB in the $\log_{10}(M_{mess})$-$\Lambda$ plane for logarithmic and
natural priors. We have also marginalized over
the messenger index $N_{mess}$. The most striking feature of the plots shown is that
while the 2D posterior with the log prior is flat in the logarithm
of the messenger scale $M_{mess}$, for the natural priors there is a very
strong peak at the highest values of $M_{mess}$. This is due to a volume
effect. With a linear or natural prior, smaller values of $M_{mess}$ occupy a
smaller prior volume, which pushes the posterior for $M_{mess}$ down. The log
priors sample more from the region of low $M_{mess}$, due to the
suppression from the log prior measure, and are in
rough agreement with the profile likelihood. The excluded region at low values
of the messenger scale in Fig.~\ref{fig:mGMSB2d}(a) is due to the requirement
that $M_{mess}>\Lambda$, since if this is not the case the messengers are
tachyonic. In order to understand why $M_{mess}$ is only very weakly
constrained by the data, consider the spectrum resulting from gauge
mediation. To leading-log order, the gaugino masses at the scale $M_{mess}$ are 
\begin{equation}
 M_{\tilde{\lambda}_i}(M_{mess}) = k_i
 N_{mess}\Lambda\frac{\alpha_i(M_{mess})}{4\pi} g(\Lambda/M_{mess}), \label{gaug}
\end{equation}
where $k_i =(5/3,1,1)$ and $k_i\alpha_i$ (no summation) are all equal at the GUT scale, and $\alpha_i$ are the gauge coupling constants. The messenger scale threshold function $g$ can be found in~\cite{hep-ph/9801271}
and $g(x)\to1$ when $\Lambda \ll M_{mess}$. This is the case throughout most of the
mGMSB parameter space that we study. In this limit one can also approximate the scalar masses by
\begin{equation}
 m^2_{\tilde{f}}(M_{mess}) = 2N_{mess}\sum_{i=1}^{3} C_i k_i
 \frac{\alpha_i^2(M_{mess})}{(4\pi)^2} \Lambda^2. \label{scal}
\end{equation}
where $C_i$ are the quadratic Casimir operators of the relevant gauge groups~\cite{hep-ph/9801271}.
The dependence on $M_{mess}$ is somewhat subtle. Increasing $M_{mess}$ leads to different
initial conditions for $\alpha_i(M_{mess})$ and also increases how far the
renormalisation group equations (RGEs) must
be evolved to reach the low scale. As pointed out in \cite{hep-ph/9609444} these
effects cancel each other out in the gaugino masses since the gaugino masses
obey the same one-loop RGEs as the gauge couplings. There is however an effect
on the scalar masses, leading to a decrease in the squark masses and an
increase in the slepton masses at larger values of $M_{mess}$. This statement
does 
not apply to the stop masses, which do depend somewhat on $\tan\beta$ due to
mixing. Therefore the mGMSB spectrum is fairly independent of $M_{mess}$.
We remind the reader that for the mGMSB fits, no DM constraint is
applied.

\begin{figure*}
\begin{center}
 \twographs{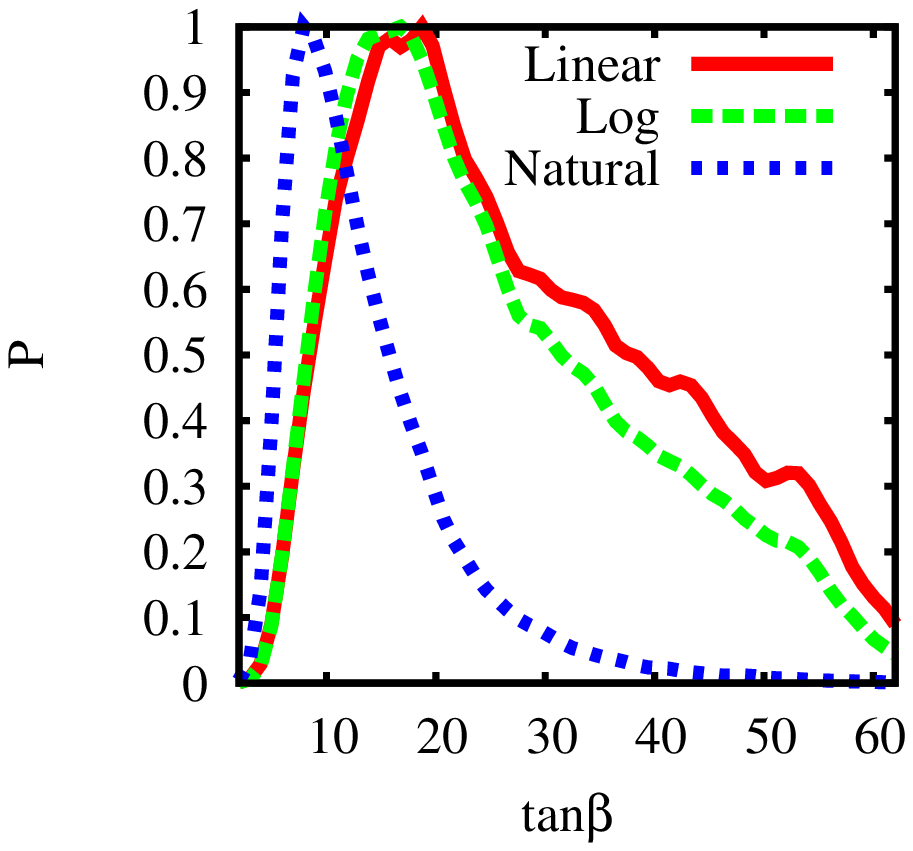}{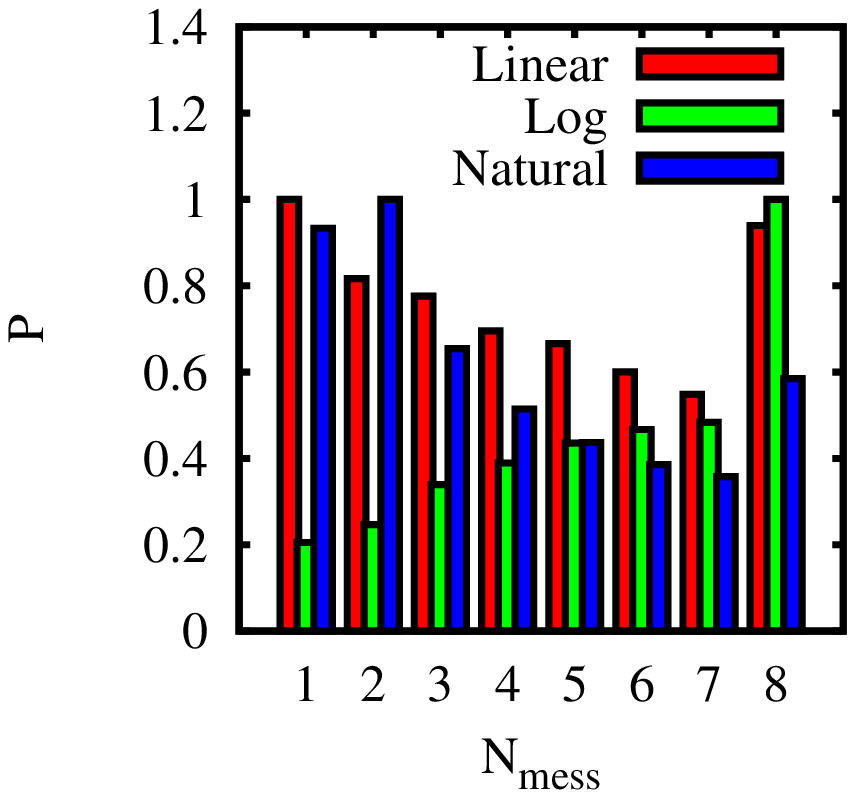}
\caption{mGMSB 1D marginalized posterior PDFs for (a) $\tan\beta$ and (b) the
  messenger index $N_{mess}$ for log, linear and natural priors.}
\label{fig:GMSBtanb}
\end{center}
\end{figure*}

Fig.~\ref{fig:mGMSB1d}(b) shows the 1D PDF posterior for the SUSY
breaking scale $\Lambda$. All sets of priors have a strong peak at around
$3\times 10^{4}$ GeV before falling away at higher values of $\Lambda$. Since
$\Lambda$ is directly related to the scale of the SUSY spectrum, high values
of $\Lambda$ lead to a heavier spectrum. Such a spectrum fits
$(g-2)_{\mu}$ and $BR(b\to s \gamma)$ worse than a lighter spectrum, as we
shall show in section~\ref{sec.discr}.

We next turn to $\tan\beta$, for which we present the posterior PDFs in
Fig.~\ref{fig:GMSBtanb}(a). The posterior PDFs for the logarithmic prior have
the same shape as the profile likelihood, with both sharply rising from the
allowed value of $\tan\beta=2$ to a most likely value just under 20 and then
gradually falling away as $\tan\beta$ increases, a feature common to both
branches of $\mu$. It is primarily the anomalous magnetic moment of the muon
that is pulling the model towards high $\tan \beta$, where a significant
SUSY component can fit the data. 
SUSY contributions to $(g-2)_\mu$ are dominated by one loop diagrams involving
chargino-sneutrino or neutralino-smuon in the loop. These are approximately
\begin{equation}
 (g-2)_{\mu}^{SUSY} \approx \frac{m_{\mu}^2 \mu \tan\beta }{16\pi^2} \left( g_1^2 F_1 M_1 + g_2^2 F_2 M_2 )\right)
\label{eq:g-2}
\end{equation}
 where $m_{\mu}$ is the muon mass and $F_{1,2}$ are positive definite
 functions of sparticle masses, which scale as $1/M_{SUSY}^4$ in the limit of
 heavy degenerate sparticles of mass $M_{SUSY}$. 
With natural priors the behaviour is similar, except that
the peak occurs at a lower value of $\tan\beta$ driven by the fact that
natural priors prefer small $\tan\beta$, which can be seen by inspecting the
Jacobian factor in Eq.~\ref{eq:natprior}. 
Fig.~\ref{fig:GMSBtanb}(b) shows the posterior PDFs 
for the discrete messenger index $N_{mess}$. 
The log priors prefer a large number of
messenger multiplets, with the most likely value of $N_{mess}$ clearly being 8. 
Since the log prior prefers light values of the neutralino and chargino
masses, a good fit to the anomalous magnetic moment of the muon is also
preferred by having relatively light scalars, so that the smuon and sneutrino 
diagrams may contribute significantly. From Eqs.~\ref{gaug} and~\ref{scal}, 
a slepton mass divided by a weak gaugino mass is proportional to
$1/\sqrt{N_{mess}}$ and so high $N_{mess}$, i.e.\ light sleptons is preferred.
This preference is less strong for the heavier spectra with the linear or
natural priors, where it is more difficult
to fit $(g-2)_\mu$. It also has to compete with volume effects coming from the
LEP2 Higgs constraint, which prefers heavy stops and therefore low $N_{mess}$.

To summarize, all of the mGMSB parameters show significant prior dependence,
even when each is marginalized down to one dimension. 
We decline to present posterior PDFs of sparticle masses, since they 
show the same non-robustness with respect to changing the priors. 

\section{mAMSB Parameter Constraints~\label{sec:fitamsb}}

In this section we discuss the effects of the observable constraints on the
anomaly mediated SUSY breaking parameter space. All supergravity
theories suffer from a Weyl (rescaling) anomaly which leads to soft breaking
terms for the visible sector. These contributions are usually rendered
negligible by the usual gravity mediated soft breaking terms. If the gravity
mediated terms are suppressed, perhaps by a mechanism similar to that
originally suggested in the brane model of \cite{Randall:1998uk} then the
SUSY breaking is dominated by the anomaly mediated terms to leading
order. In pure AMSB the slepton masses suffer from being tachyonic. One way
this can be ameliorated while avoiding the SUSY flavour problem is by
introducing a new universal mass parameter $m_0$ at the GUT scale which lifts
the slepton masses. We define the GUT scale to be the scale of electroweak
gauge unification.
The GUT-scale soft breaking terms for the gauginos,
scalars and trilinear couplings are then given by 
\begin{equation}
 M_{i} = \frac{\beta_{g_i}}{g_i}m_{3/2}\label{eq:AMSB1},
\end{equation}
\begin{equation}
 m^2_{\tilde{Q}} = -\frac{1}{4} \left( \frac{\partial\gamma}{\partial g_i}\beta_{g_i} + \frac{\partial\gamma}{\partial y}\beta_y \right) m_{3/2}^2 + m_0^2,
\label{eq:AMSBslep}
\end{equation}
\begin{equation}
 A_y = - \frac{\beta_y}{y} m_{3/2}, \label{eq:AMSB3}
\end{equation}
where $\beta_{g_i}$ is the beta function of the $i^{th}$ SM gauge group,
$\gamma$ is the anomalous dimension of the respective scalar wave-function and
$y$ and $\beta_y$  are Yukawa couplings and Yukawa beta functions respectively. Eqs.~\ref{eq:AMSB1}-\ref{eq:AMSB3} constitute the
``minimal AMSB'' (mAMSB) SUSY breaking assumption.
Its spectrum is specified by three
continuous free parameters: the gravitino mass $m_{3/2}$, the parameter $m_0$
and $\tan\beta$, along with the sign of the $\mu$ parameter. Our mAMSB fits
use the `asymmetric' DM constraint shown in Fig.~\ref{fig:omega}, since the mAMSB neutralino relic density is too low to be the sole component of the dark matter.

\begin{figure}
\begin{center}
\sixgraphs{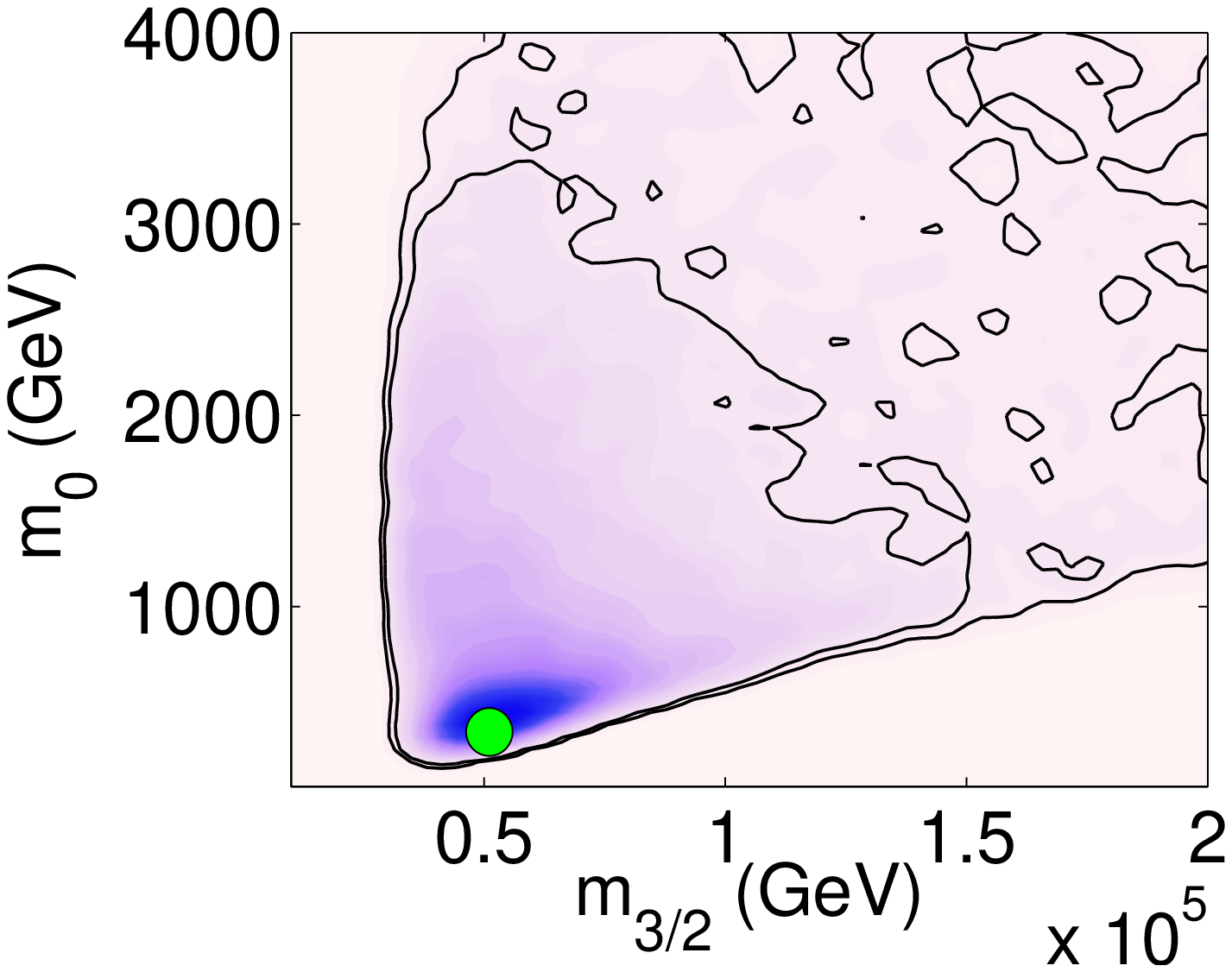}{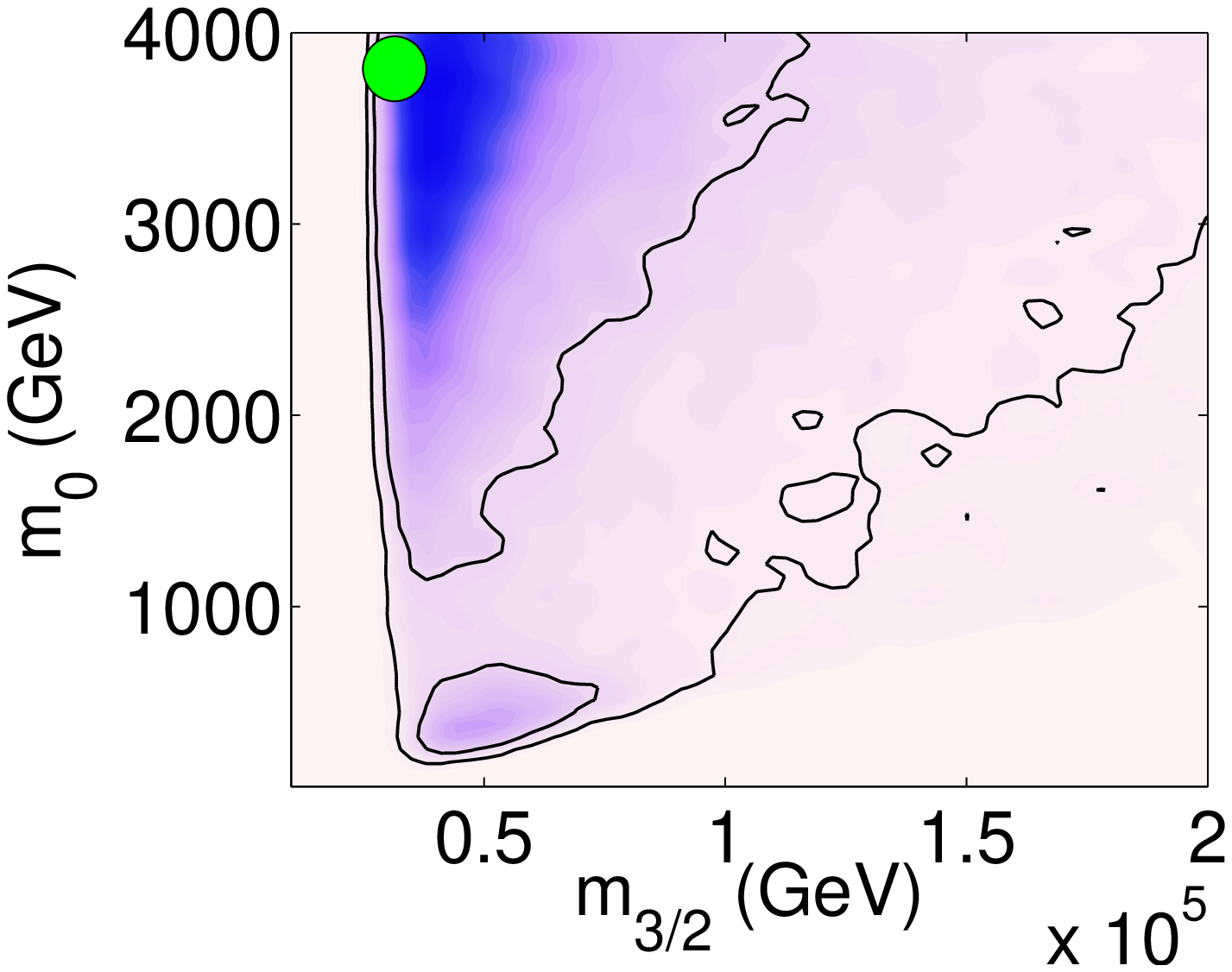}{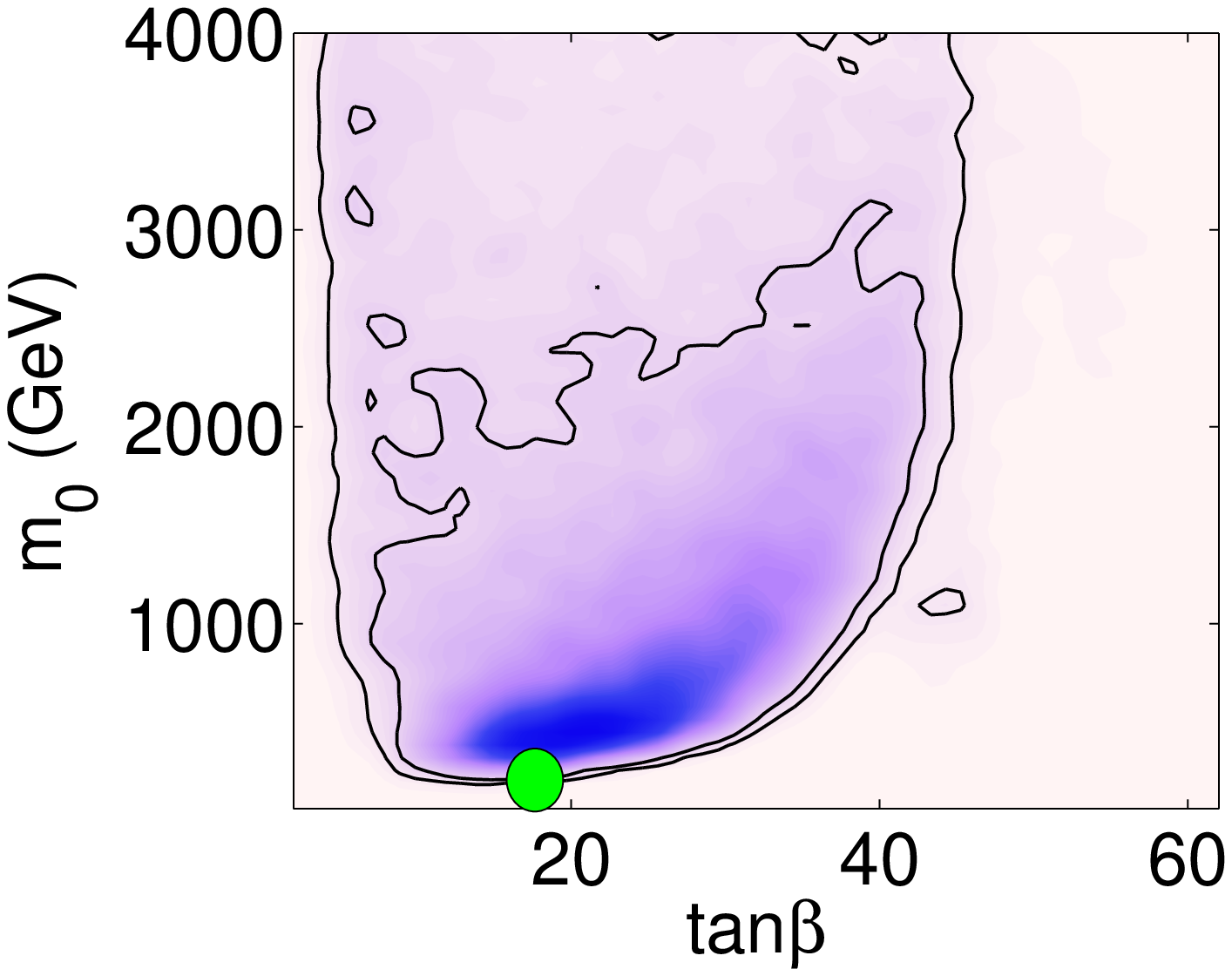}{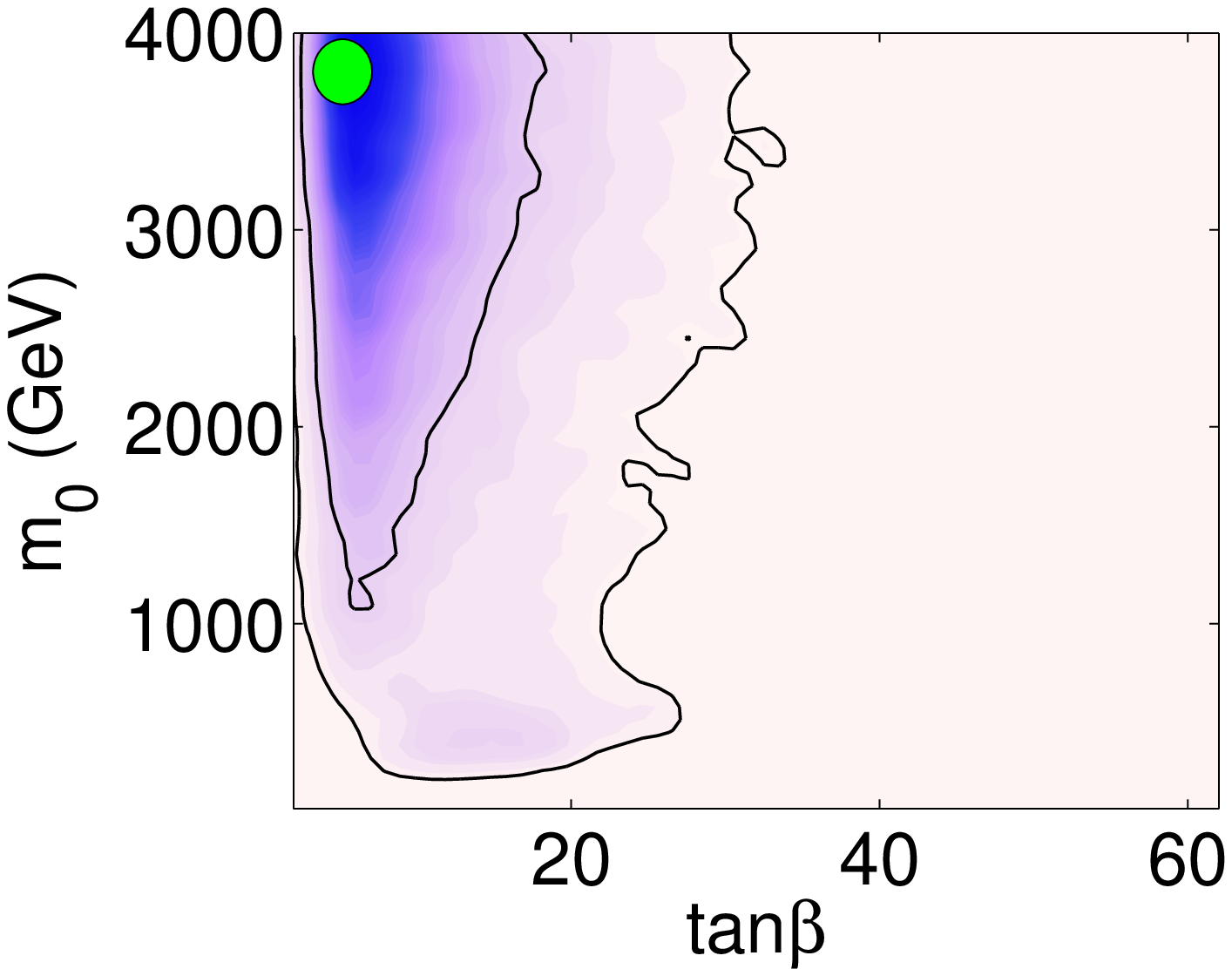}{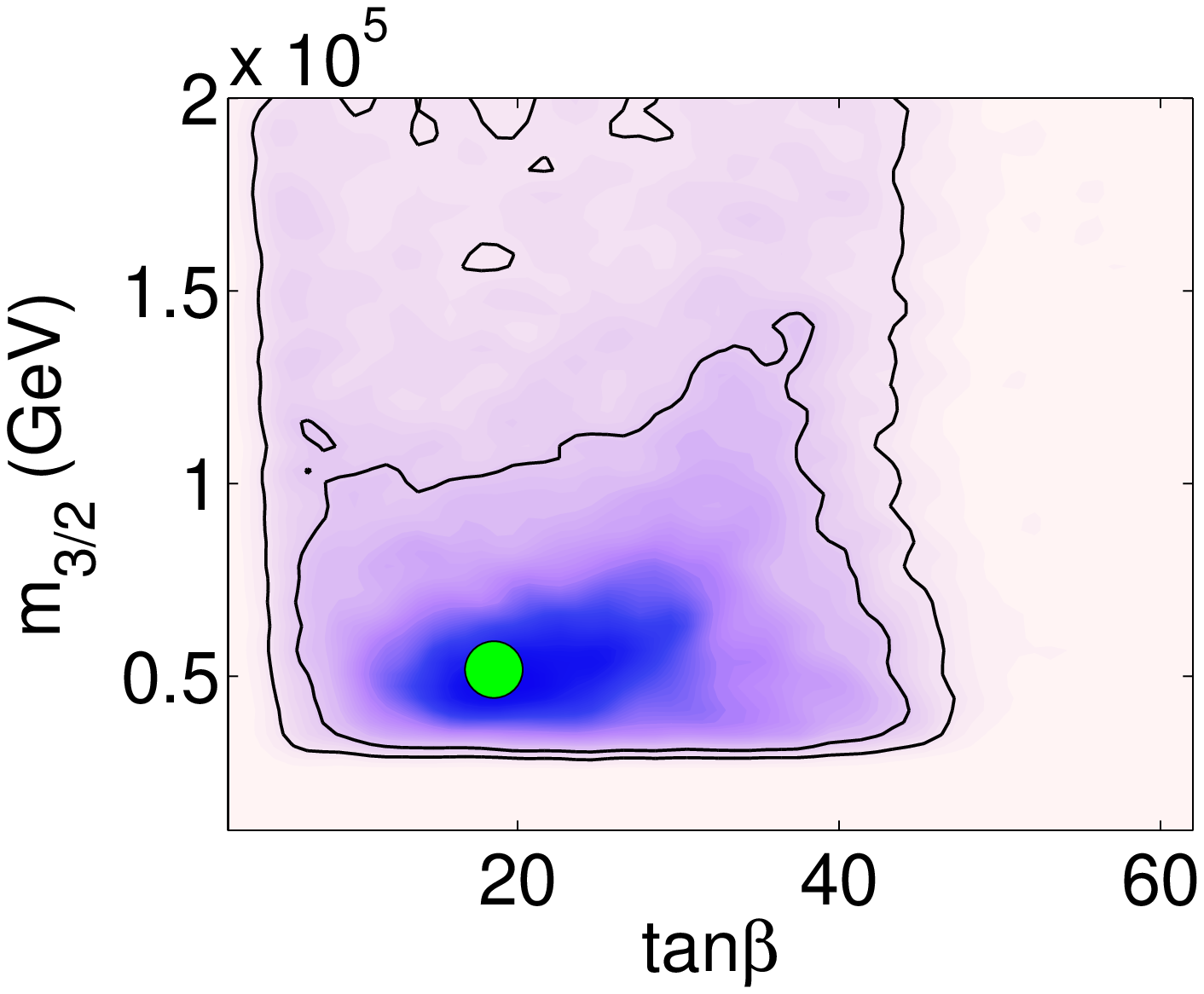}{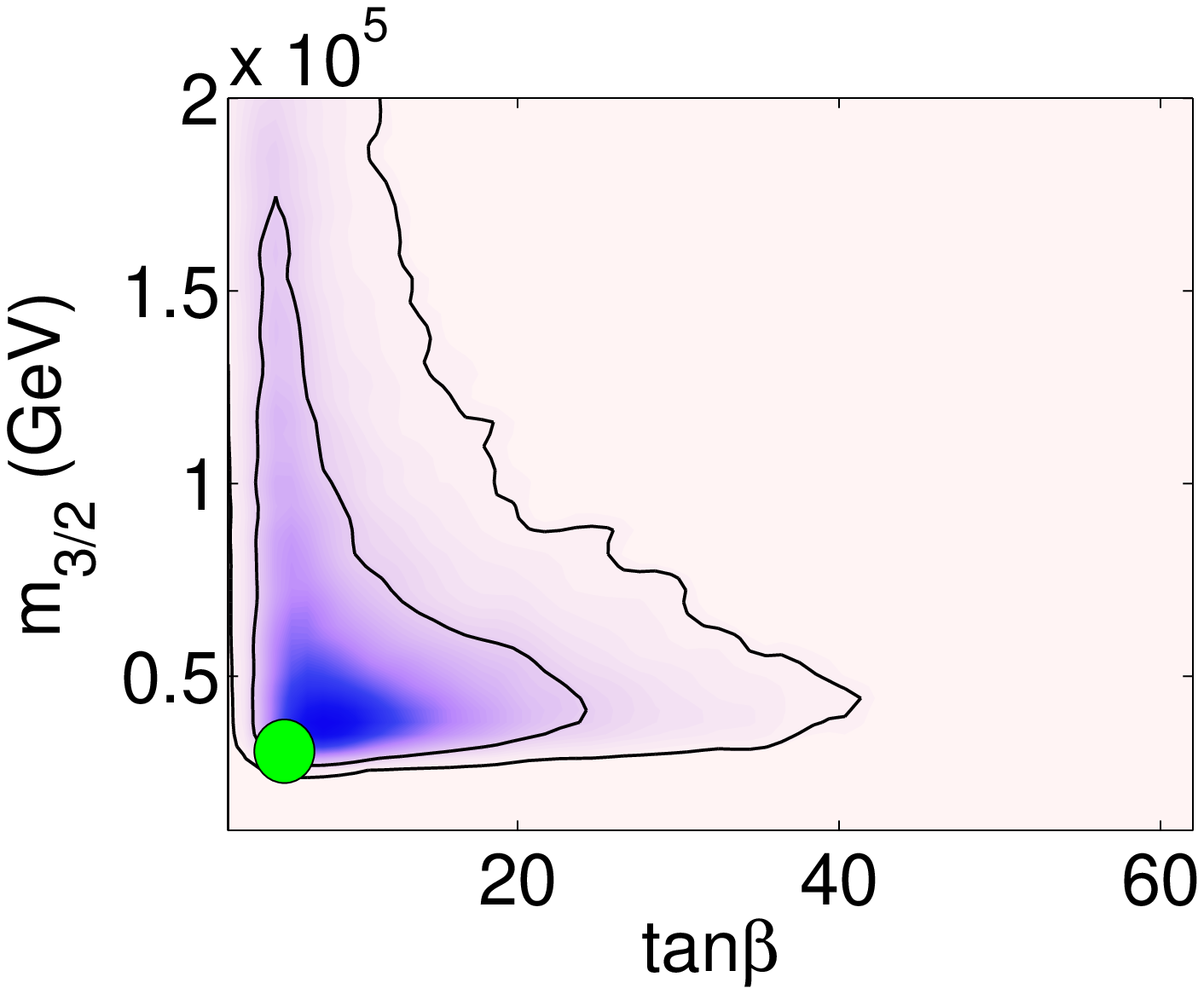}
\caption{mAMSB marginalised 2D posterior PDFs in the $m_0$-$m_{3/2}$,
  $m_0$-$\tan\beta$ and $m_{3/2}$-$\tan\beta$ planes for linear priors
  (left-hand side pictures) and natural priors (right-hand side
  pictures) with asymmetric $\mathcal{L}_{DM}$. The green dots mark the best-fit points in each prior case sample.
  68$\%$ and 95$\%$ Bayesian credibility regions are shown as the
  inner and outer contours respectively.}
\label{fig:AMSB2D}
\end{center}
\end{figure}

Fig.~\ref{fig:AMSB2D}(a,c,e) shows the posterior PDFs for linear priors in the
$m_{3/2}$-$m_0$, $m_{3/2}$-$\tan\beta$ and $m_0$-$\tan\beta$ planes while
Fig.~\ref{fig:AMSB2D}(b,d,f) shows the same plots for natural priors. The dark matter constraint is asymmetric in both cases. There is
clearly a strong prior dependence in all parameters. This is due to the
pull of the natural priors to low values of $\tan\beta$ and to high values of
$B/\mu$, which occurs near the focus point at high $m_0$~\cite{Feng:1999zg}.
The 
linear priors identify a different region of high probability at low $m_0$ and
low $m_{3/2}$ leading to a light sparticle spectrum capable of satisfying the
constraints from $(g-2)_{\mu}$ and $BR(B\to X_s\gamma)$. Log prior posteriors are
omitted for brevity, but they are
similar to the linear prior results, except that the log priors have an even
stronger preference for low $m_0$ and $m_{3/2}$. 

There are three significant regions of parameter space which have been
ruled out already. The first is the triangular region at low $m_0$ and
moderate to high values of $m_{3/2}$. This is ruled out by the direct
search constraints on the sleptons. The large values of $m_{3/2}$ and
small $m_0$ lead to low (or even imaginary) values of the slepton
masses in Eq.~\ref{eq:AMSBslep}. The disallowed strip at low values of
$m_{3/2}$ and all values of $m_0$ is due to gaugino masses being too
low. The region defined by $\tan\beta\gsim50$ is decisively disfavoured
by a combination of related circumstances: tachyons, no radiative
electroweak symmetry breaking and, as $m_0$ gets larger, a Higgs
potential that is unbounded from below. 

\begin{figure}
\begin{center}
\AMSBthreegraphs{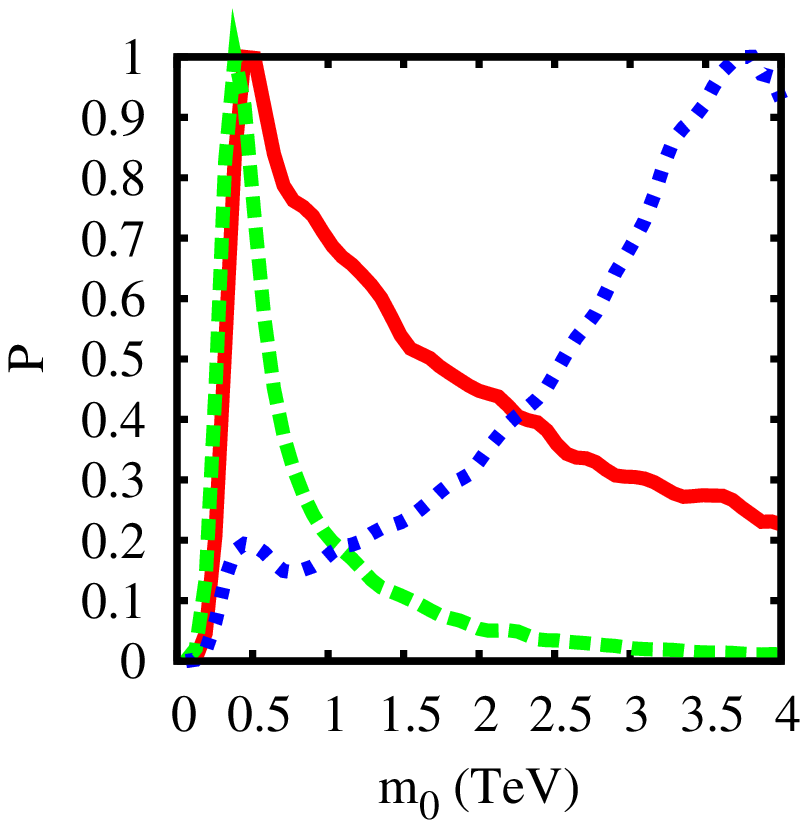}{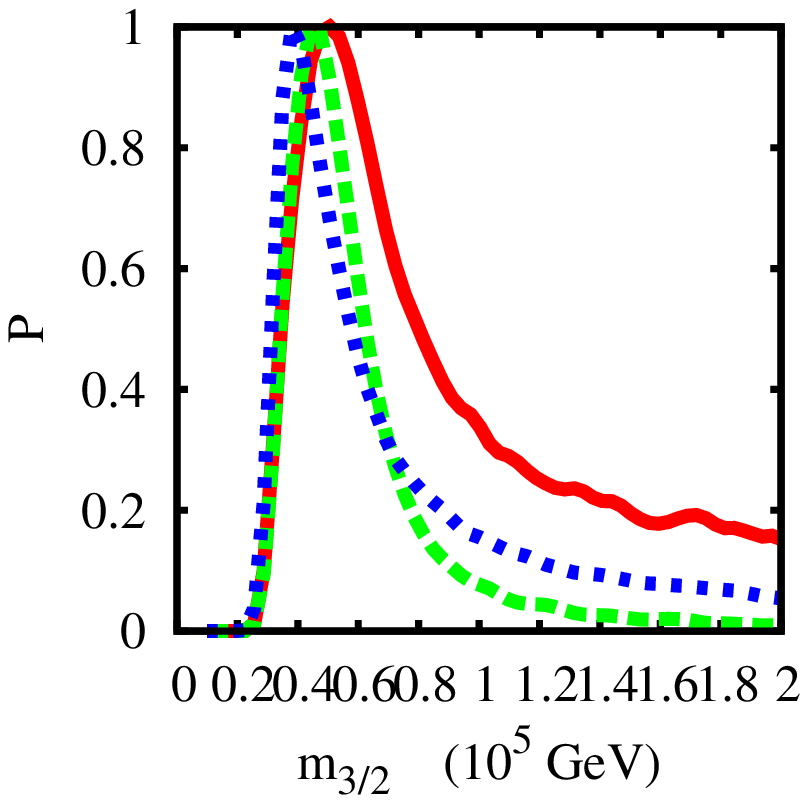}{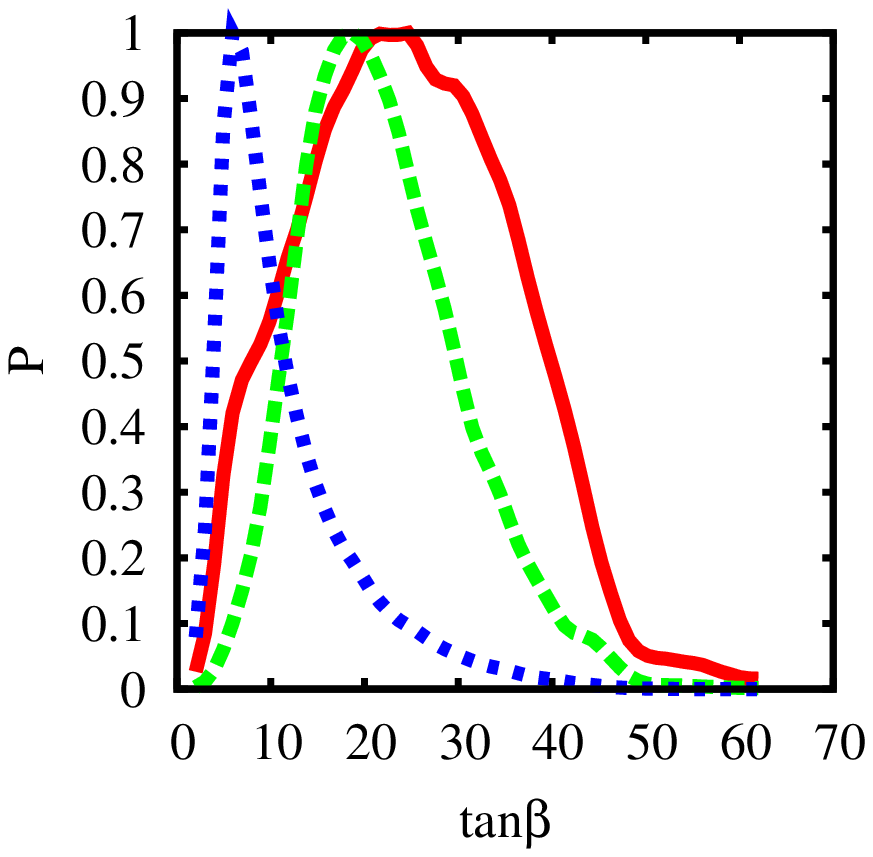}
\caption{mAMSB marginalized 1D
  posterior PDFs for 
(a) $m_0$, (b) $m_{3/2}$ and (c)
  $\tan\beta$ for different priors. Red (solid) lines represent linear priors,
  green (dashed) lines for log priors and blue (dotted) lines for natural
  priors. All plots are for asymmetric $\mathcal{L}_{DM}$.} 
\label{fig:AMSB1D}
\end{center}
\end{figure}

Fig.~\ref{fig:AMSB1D} shows the 1D posterior PDFs for (a) $m_0$,
(b) $m_{3/2}$ and (c) $\tan\beta$ for all three priors used with
red (solid line) representing the linear priors, green (dashed lines)
the log priors and blue (dotted lines) the natural
priors. The dark matter constraint is asymmetric in all cases. As expected the logarithmic priors show a distinct preference for low
$m_0$, with the linear priors expressing the same behaviour except for a longer
and less suppressed tail to high $m_0$. The natural priors have the
opposite behaviour, indicating that the fits are strongly prior
dominated. Despite 
the aforementioned statistical noise, profile likelihoods do indicate that
the region of low $m_0$ favoured by the linear and log priors is a better fit
to the data than the region favoured by the natural priors. Indeed, part of this
peak is likely due to a volume effect due to the large probability mass at
high $m_0$. 
There is much better agreement between the three sets of priors for
the posterior PDF for $m_{3/2}$ with all priors exhibiting a peak in
the likelihood at near 50 TeV falling off quite rapidly in the high
$m_{3/2}$ region. The posterior PDF for $\tan\beta$ shows two
different peaks. One of these is due to the natural priors and is at
low $\tan\beta\sim 8$ due to the natural priors preference for low
values of this variable. The other, broader peak is a better fit and
occurs around $\tan\beta\sim22$. This moderate value of $\tan\beta$ is
preferred due to the statistical pull from observables such as
$(g-2)_{\mu}$, $BR(B\to X_s\gamma)$ and $BR(B_s\to \mu^+\mu^-)$, and
values higher than this are disfavoured due to the SM constraints on
these same observables as is discussed in more detail
in~\cite{arXiv:0902.4880}.   

To summarize, like mGMSB, all of the mAMSB parameters show significant prior
dependence, even when each is marginalized down to one dimension. 
We decline to present posterior PDFs of sparticle masses, since they 
show the same non-robustness with respect to changing the priors. We turn now to a more detailed examination
of the constraints from the observables.

\section{Constraining Power and Statistical Pull of Observables \label{sec:con}}

\subsection{Information Content and Constraining Power}\label{sec:results:KL}

In order to calculate the information gain in moving from a prior distribution to the posterior distribution, one
can calculate the Kullback-Leibler (KL)~\cite{KLdiv} divergence (also called information divergence, information gain,
or relative entropy) between the prior and the posterior distributions. Denoting the posterior distribution and
prior distributions by $p(\underline m|\underline d, H)$ and $p(\underline
m | H)$ respectively, the KL divergence of observable $i$
is defined as
\begin{equation}
D_{KL}=\int p(\underline m
| \underline d, H) \log \left( \frac{p(\underline m | \underline d, H)}{p(\underline
  m)} \right). \label{eq:KL1}
\end{equation} $D_{KL}$ quantifies how much the prior PDF has been updated in
its 
transformation to the posterior PDF. 
Using Eqs.~\ref{eq:KL1},\ref{eq:bayes}, the KL divergence of observable $i$ is
\begin{equation}
D_{{\rm KL}_i} = \int \frac{{\mathcal L}_i(\underline m) p(\underline m | H)}{p(d_i | H)}
\log \left[ \frac{\mathcal{L}_i(\underline m)}{p(d_i | H)} \right] d\underline m,
\label{eq:D_KL}
\end{equation}
where $p(d_i|H)=\int \mathcal{L}_i(\underline m) p({\underline m} | H) d {\underline m}$.
Since the Standard Model observables are used in our fits as inputs, we consider them to be a part of the prior. 
The KL divergence of all of the data combined $D_{{\rm KL}}$ is easily
obtained from the {\sc MultiNest} algorithm  
which already calculates it  to obtain the uncertainty on the evidence
estimate (see~\cite{Feroz:2007kg}). We defined the constraining power of observable
$i$ to be
\begin{equation}
C_{P_i} = \frac{D_{{\rm KL}_i}}{D_{{\rm KL}}}. 
\end{equation} 
The KL divergence was also employed in~\cite{Trotta:2008bp} to calculate the constraining power of observables 
for the $\mu > 0$ branch of the CMSSM using linear and log priors and it was
concluded that the information gain 
is  dominated by the DM constraint which alone accounts for $C_P=0.8-0.95$ depending on the prior. By examining
the variation of the $95\%$ CL  regions in the CMSSM parameter space with and without the DM
constraint,~\cite{Buchmueller:2008qe} on the other hand concluded that the DM has little role in constraining the
parameter space of the CMSSM\@.  We update the analysis of~\cite{Trotta:2008bp} by including the additional
electroweak observables returned by \texttt{SusyPOPE} and calculating $D_{\rm
  KL}$ for the observables for the CMSSM
with both signs of $\mu$. We list the $C_{P_i}$ values in 
Table~\ref{tab:KL} for the CMSSM, mAMSB and LVS respectively. 
\begin{table}
\begin{center}
\begin{tabular}{|c|c|c|c|c|c|c|}
\hline
$C_{P_i}$& \multicolumn{3}{c|}{symmetric $\mathcal{L}_{\rm DM}$} &
\multicolumn{3}{c|}{asymmetric $\mathcal{L}_{\rm DM}$} \\ \hline
Observables/Prior  & linear & log & natural & linear & log & natural \\ \hline \hline
 & \multicolumn{6}{|c|}{CMSSM} \\ \hline
$\Omega_{\rm DM}h^2$  			& $0.65$ & $0.57$ & $0.63$ & $0.56$ & $0.47$ & $0.56$ \\ 
B-Physics    				& $0.44$ & $0.40$ & $0.41$ & $0.49$ & $0.45$ & $0.48$ \\ 
$BR(B\to X_s \gamma)$    		& $0.43$ & $0.40$ & $0.40$ & $0.48$ & $0.44$ & $0.47$ \\ 
Electroweak    				& $0.49$ & $0.36$ & $0.45$ & $0.53$ & $0.38$ & $0.55$ \\ 
$\delta a_{\mu}$  			& $0.41$ & $0.32$ & $0.37$ & $0.44$ & $0.34$ & $0.48$ \\ 
$m_h$  					& $0.42$ & $0.39$ & $0.39$ & $0.46$ & $0.42$ & $0.47$ \\ 
\hline\hline
 & \multicolumn{6}{|c|}{mAMSB} \\ \hline
$\Omega_{\rm DM}h^2$  			& $0.02$ & $0.01$ & $0.01$ & $0.01$ & $0.00$ & $0.00$ \\ 
B-Physics    				& $0.03$ & $0.06$ & $0.03$ & $0.04$ & $0.07$ & $0.03$ \\ 
$BR(B\to X_s \gamma)$    		& $0.02$ & $0.05$ & $0.01$ & $0.03$ & $0.05$ & $0.02$ \\ 
Electroweak    				& $0.12$ & $0.11$ & $0.14$ & $0.13$ & $0.11$ & $0.15$ \\ 
$\delta a_{\mu}$  			& $0.01$ & $0.05$ & $0.01$ & $0.02$ & $0.06$ & $0.02$ \\ 
$m_h$  					& $0.02$ & $0.05$ & $0.04$ & $0.03$ & $0.05$ & $0.05$ \\ 
\hline\hline
 & \multicolumn{6}{|c|}{LVS} \\ \hline
$\Omega_{\rm DM}h^2$  			& $0.49$ & $0.45$ & $0.47$ & $0.26$ & $0.16$ & $0.23$ \\ 
B-Physics    				& $0.19$ & $0.23$ & $0.18$ & $0.25$ & $0.27$ & $0.22$ \\ 
$BR(B\to X_s \gamma)$    		& $0.19$ & $0.22$ & $0.17$ & $0.24$ & $0.27$ & $0.23$ \\ 
Electroweak    				& $0.14$ & $0.11$ & $0.13$ & $0.15$ & $0.11$ & $0.14$ \\ 
$\delta a_{\mu}$  			& $0.09$ & $0.05$ & $0.08$ & $0.12$ & $0.06$ & $0.11$ \\ 
$m_h$  					& $0.20$ & $0.20$ & $0.21$ & $0.25$ & $0.25$ & $0.26$ \\ 
\hline
\end{tabular}
\caption{Constraining power of different observables in the CMSSM, mAMSB and LVS
  marginalized over both signs of $\mu$. 
`B-Physics' includes the
observables in the third section of Table~\ref{tab:obs}. Electroweak observables include the ones in the second
section of Table~\ref{tab:obs} plus $m_W$ and $\sin^2 \theta^l_{eff}$. Symmetric $\mathcal{L}_{\rm DM}$ denotes
the assumption that the DM relic density is composed entirely of the LSP and asymmetric $\mathcal{L}_{\rm DM}$
codifies the assumption that the LSP forms only a part of the DM relic density.}
\label{tab:KL}
\end{center}
\end{table}

Table~\ref{tab:KL} shows that in the CMSSM, the DM constraint is dominant
in constraining the parameter space, as is familiar from current
literature. Since the asymmetric $L_{DM}$ rules out less of parameter space than the symmetric constraint, $C_{P_{\Omega_{DM}h^2}}$ is always smaller in the asymmetric case.
However,
$B-$physics constraints
dominated by $BR(B \rightarrow X_s 
\gamma)$, electroweak constraints, the anomalous magnetic moment of the muon
and the Higgs mass constraint all play significant roles too. 
By contrast,
mAMSB, which predicts approximately zero DM relic density over its
entire parameter space, is hardly constrained at all by DM constraints. In
fact, none of the indirect constraints provide much constraining power except
for the electroweak observables. Once one has taken the SM inputs into account, the indirect observables add very little to the degree of constraint on parameter space.
The LVS is most highly constrained by DM, although less so than the CMSSM. 
This is because~\cite{Allanach:2008tu} a larger volume of the prior
corresponds to a relic density compatible with the WMAP DM constraint compared
to the CMSSM\@.  All of the other indirect observables also help to constrain
the LVS scenario in a non-trivial way. 
Defining $C_{SM}$ where $SM = \{ \alpha_s(M_Z)^{\overline{MS}},
\alpha(M_Z)^{\overline{MS}},  m_t, m_b, m_Z \}$, we consistently find values
$C_{SM}=$0.8--0.9 for each of the three models investigated in Table~\ref{tab:KL},
regardless of the form of the DM constraint or prior. It is therefore
essential to vary these input parameters and to use the available experimental
data to constrain them.

\subsection{Statistical Pull of Observables}\label{sec.gstatsp}

\subsubsection{mGMSB}
\begin{figure}
\begin{center}
\sevengraphs{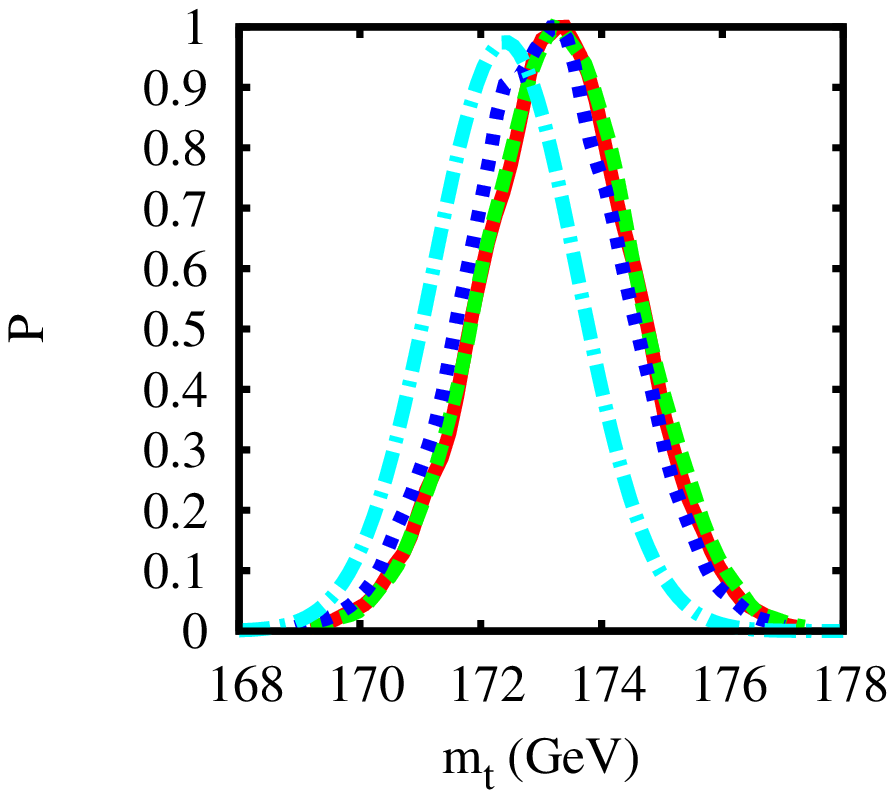}{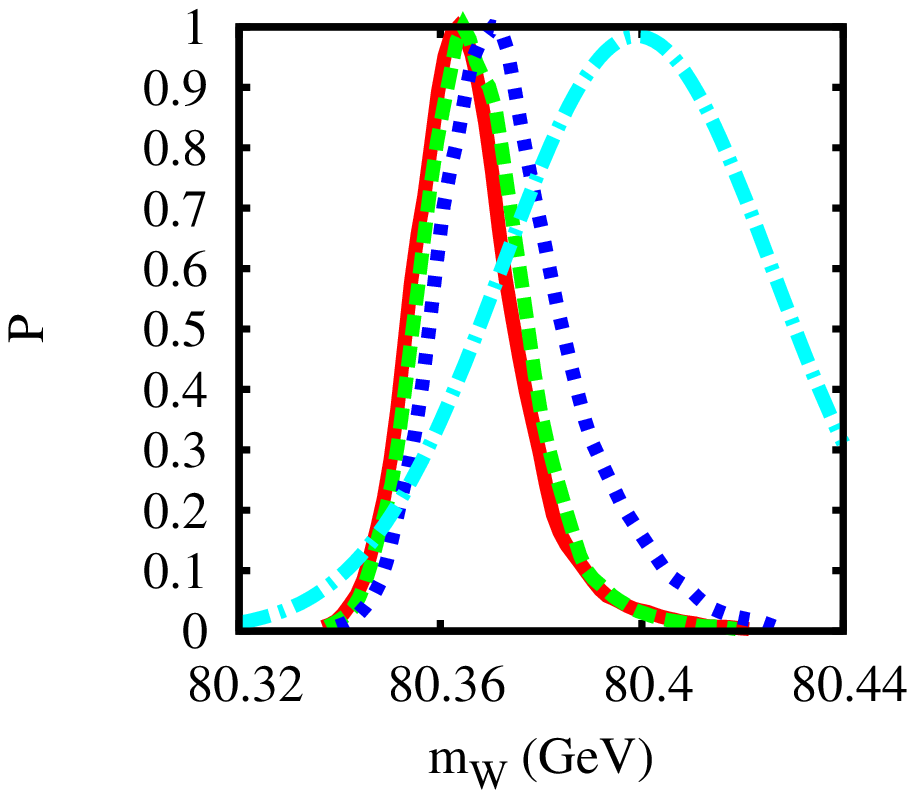}{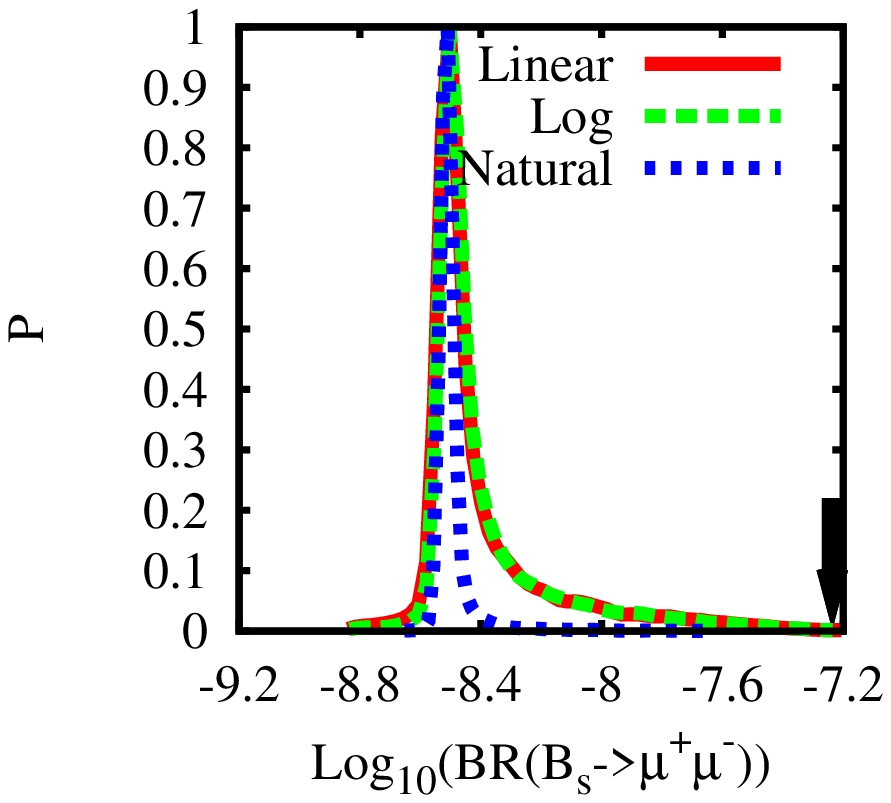}{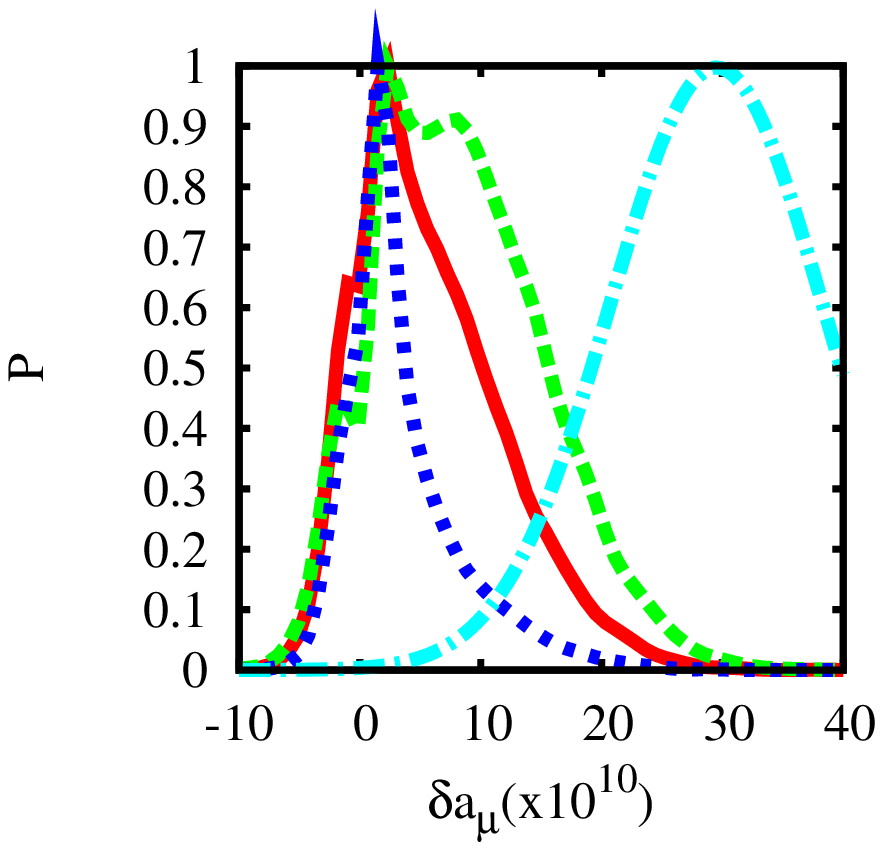}{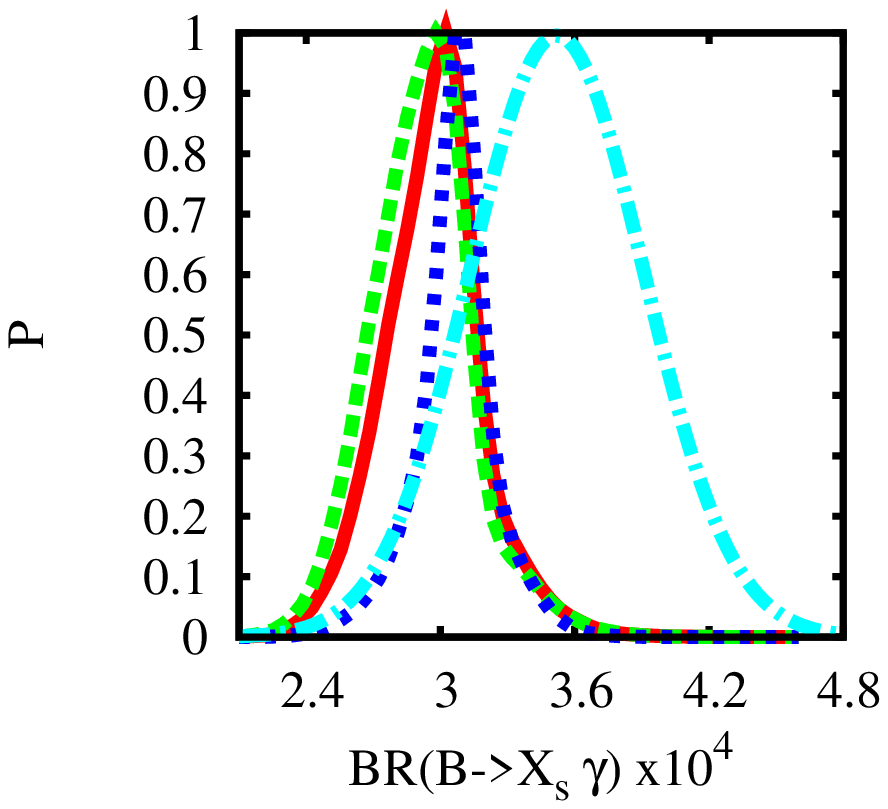}{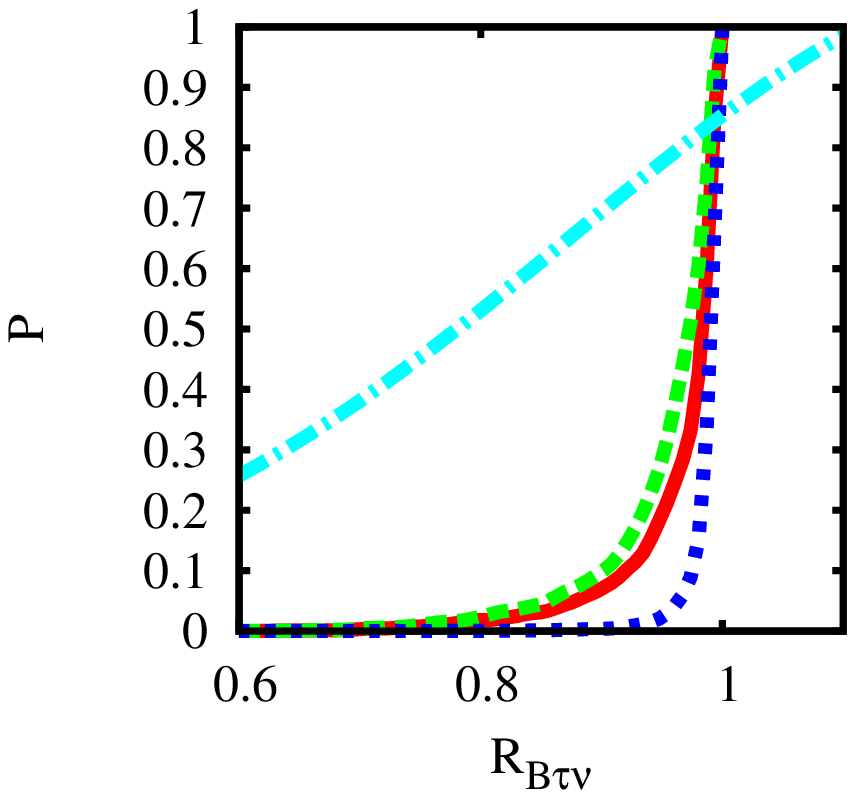}{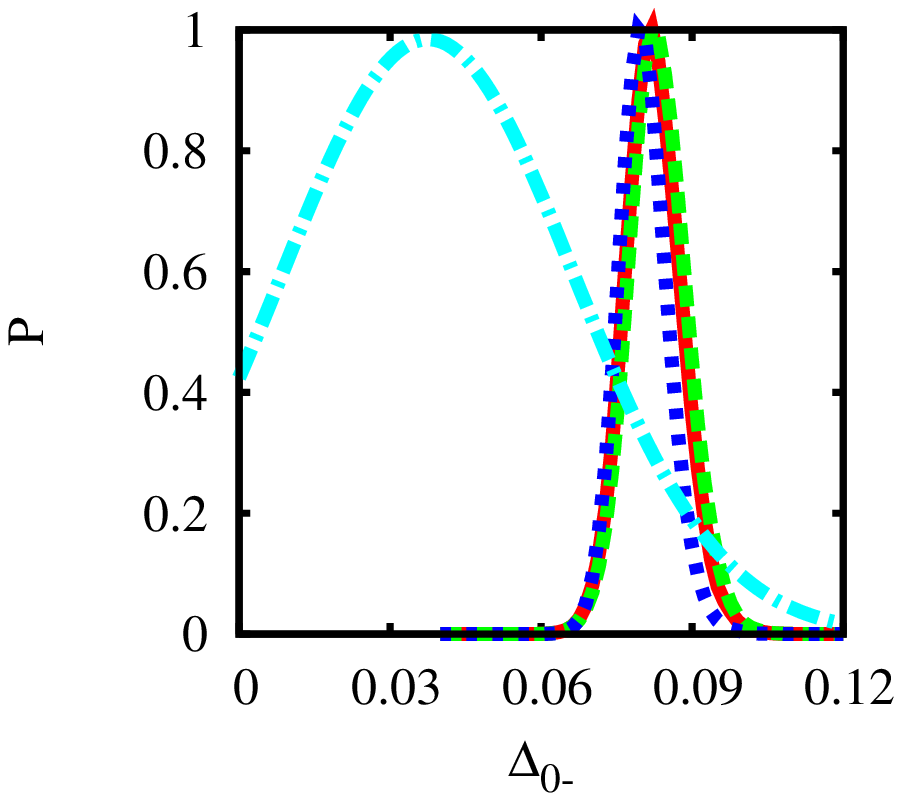}
\caption{The statistical pull of some important constraints on
  mGMSB, where no dark matter constraint is applied. Posterior PDFs are plotted for: 
red (solid) lines have linear priors,
green (dashed) lines have log priors, blue (dotted) lines have natural
priors and cyan (dash-dotted) is the experimental likelihood
constraint. The figures show (a) the top mass $m_t$, (b) the W mass $m_W$, (c)
the logarithm of the branching ratio for $B_s\to \mu^+ \mu^-$ with the black
arrow being the current  
95$\%$ CL experimental upper limit, (d) the anomalous magnetic moment
$\delta a_{\mu}$, (e) $BR(B\to X_s\gamma)$, (f) the branching ratio 
$BR(B_u \rightarrow \tau\nu)$ divided by its SM prediction and (g) the isospin
asymmetry in $B \rightarrow K^* \gamma$ decays $\Delta_{0-}$.} 
\label{fig:GMSBpull}
\end{center}
\end{figure}

We now examine the pull exerted by the experimental constraints. Some of the
more important of these are shown in Fig.~\ref{fig:GMSBpull}.
Considering first the posterior PDF for $m_t$, there is rough equality between
the experimental constraint and that predicted by mGMSB, indicating that it is
not being strongly pulled by other observables. 

The W boson mass is typically under predicted by about $1\sigma$ as compared to
current direct measurements of its mass. The fits show
a weak correlation between $m_t$ and $m_W$ such that a 
decrease in the experimental value of $m_W$ would also lead to a better fit
for $m_t$. This effect is quite small however, as is the overall disagreement
between theory and experiment. The PDFs for $m_b$ closely follows its
experimental constraint and 
$\sin^2\theta^{l}_{eff}$ is constrained to be close to its central
value. These two observables are not shown, for brevity.
In Fig.~\ref{fig:GMSBpull}(c) we show the logarithm of the branching ratio
for the flavour changing decay $B_s\to\mu^+\mu^-$. 
The 95$\%$ CL experimental upper limit from the Tevatron is
marked with a black arrow at the right of the plot. 
If this branching ratio is measured to be non-zero in the near future, it
will be far above the SM prediction and therefore provide a strong constraint
on mGMSB, removing most of the currently available posterior density, which
resides near the SM prediction of $\sim 10^{-8.5}$.

The $3.2\sigma$ discrepancy between observation and the Standard Model
theoretical value of the anomalous magnetic moment of the muon has received
much attention from SUSY phenomenologists. In mGMSB the sign of the
SUSY corrections to this quantity are correlated with the sign of
$\mu$. The $\mu>0$ branch thus fits the data much better than the $\mu<0$
branch. For $\mu>0$ the natural priors prefer small values of $\delta a_{\mu}$,
since the SUSY contributions to $\delta a_{\mu}$ are proportional to
$\tan\beta$, large values of which are suppressed by the natural prior. 
Log priors, while still exhibiting a preference for low values of
$\delta a_{\mu}$, can provide a better fit.

Fig.~\ref{fig:GMSBpull}(e) shows that mGMSB predicts the most likely branching
ratio of $B\to X_s \gamma$ to be between 1$\sigma$ less than the
experimental central value. 
The stop-chargino contribution interferes (for positive
values of $M_3$ and $A_t$ such as we have in mGMSB) opposite to the
sign of $\mu$ with respect to the SM contribution. When $\mu>0$ as
preferred by $\delta a_\mu$, it destructively interferes.
This is mitigated somewhat by the charged
Higgs-top contribution, which always interferes constructively with the SM
matrix element. 
The interplay between these two dominant SUSY contributions
still leads to a predicted value below the observed one in mGMSB\@. For natural
priors 
the PDF is quite sharply peaked around the SM value of $3.15\times
10^{-4}$. For log and linear priors the peak is shifted slightly to
lower values, with the tails of the distributions being longer than
that of the natural priors. 

Fig.~\ref{fig:GMSBpull}(f) shows the posterior PDF of the ratio
$R_{B\tau\nu}\equiv 
BR(B_u\to\tau\nu)/BR(B_u\to\tau\nu)_{SM}$, where $BR(B_u\to\tau\nu)_{SM}$ is
the SM prediction of the branching ratio. The expressions for the SUSY
corrections to this quantity can be found in \cite{arXiv:0710.2067, arXiv:0808.3144}. The SUSY
contribution is always negative, so that the total MSSM prediction is smaller
than that predicted by the Standard Model. The experimental value of
$R^{exp}_{B\tau\nu}=1.28\pm0.40$ is compatible with a non-zero SUSY
contribution. However, a more precise experimental measurement of this value
which had the same central value but half the uncertainty would be better fit
without SUSY, and could lead to some tension with the $(g-2)_{\mu}$
measurement.   
The isospin asymmetry in $B \rightarrow K^* \gamma$ decays $\Delta_{0-}$ is
shown in Fig.~\ref{fig:GMSBpull}(g). mGMSB prefers a larger asymmetry than is
observed. 
The discrepancy between the mode value and the experimental central value is
just under $2\sigma$. 

We have also investigated the PDFs for the electroweak
observables calculated using \texttt{SusyPope}. We have found that they do not
exhibit enough variation over the parameter space to present large constraints
on our fits. In particular $R_{l}^0$, $R_c^0$, $R_b^0$, $\mathcal{A}_b$, $\mathcal{A}_c$ and
$A_{fb}^{0,c}$ are all consistent with experiment independently of the
prior. The forward backward asymmetry $A_{fb}^{0,b}$ is predicted too large by
2.4 $\sigma$ while the left-right asymmetry $A_{LR}^0$ is too small by 2
$\sigma$.

To summarize the mGMSB parameter space fits, we observe a similar significant
level of prior dependence in mGMSB 
to that previously observed in the 
CMSSM~\cite{Allanach:2005kz,Allanach:2006jc,arXiv:0705.0487,hep-ph/0609295,Feroz:2008wr,Trotta:2008bp,Feroz:2009dv}. It
is clear from 
Figs.~\ref{fig:mGMSB1d} and~\ref{fig:GMSBtanb} that none of the parameters are
constrained independently of prior and so the fits require additional more
precise and/or direct data before they can be considered robust. 

\subsubsection{mAMSB}

\begin{figure}
\begin{center}
 \sevengraphs{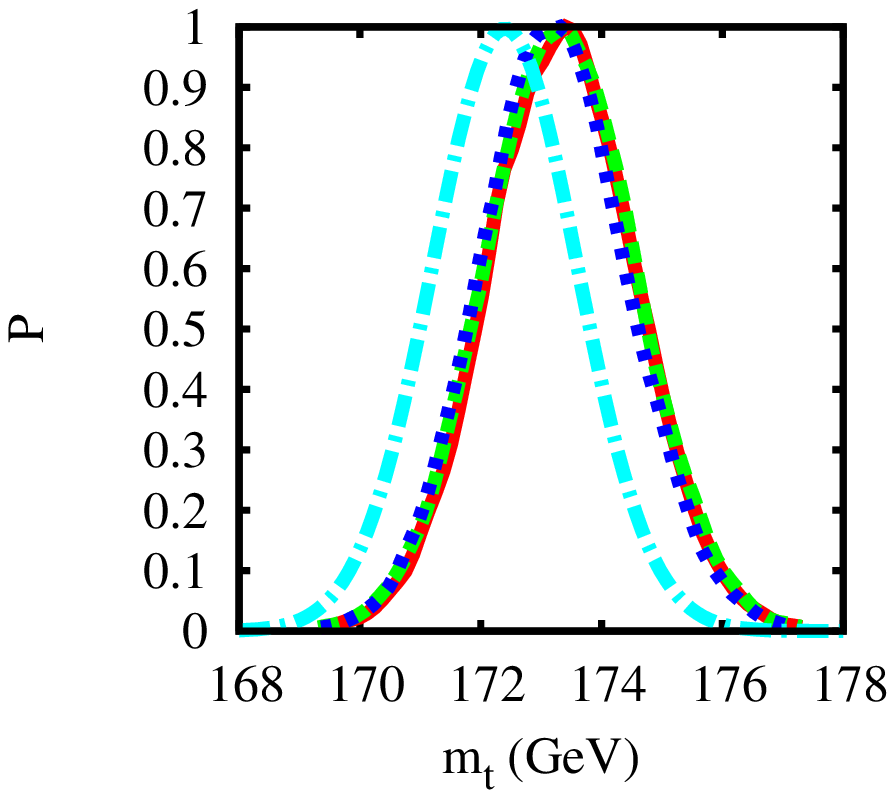}{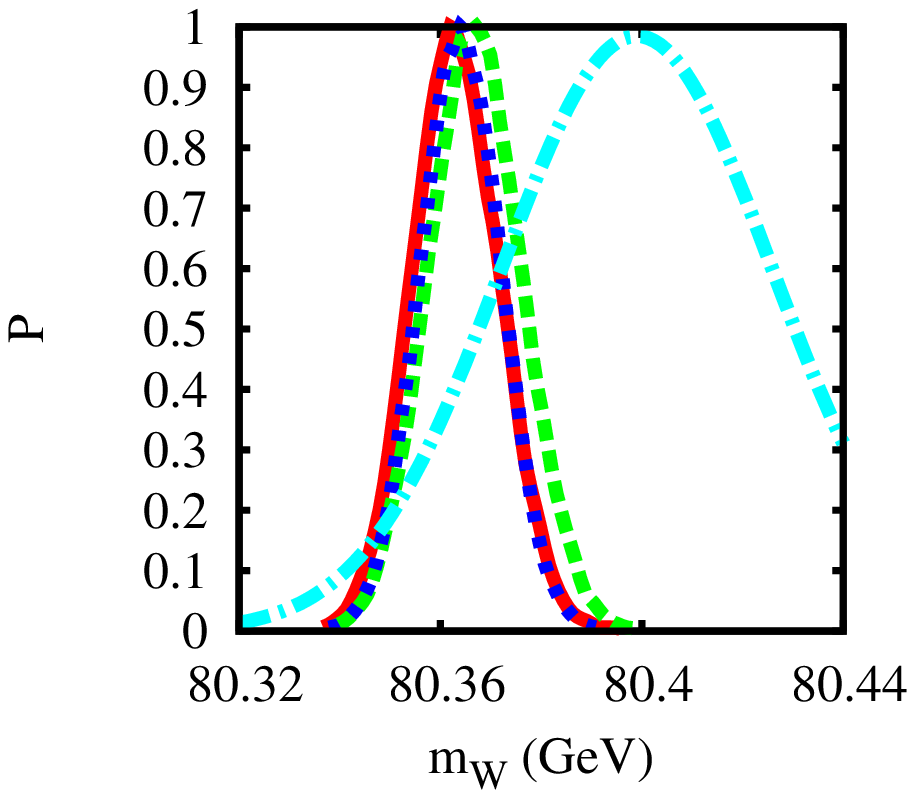}{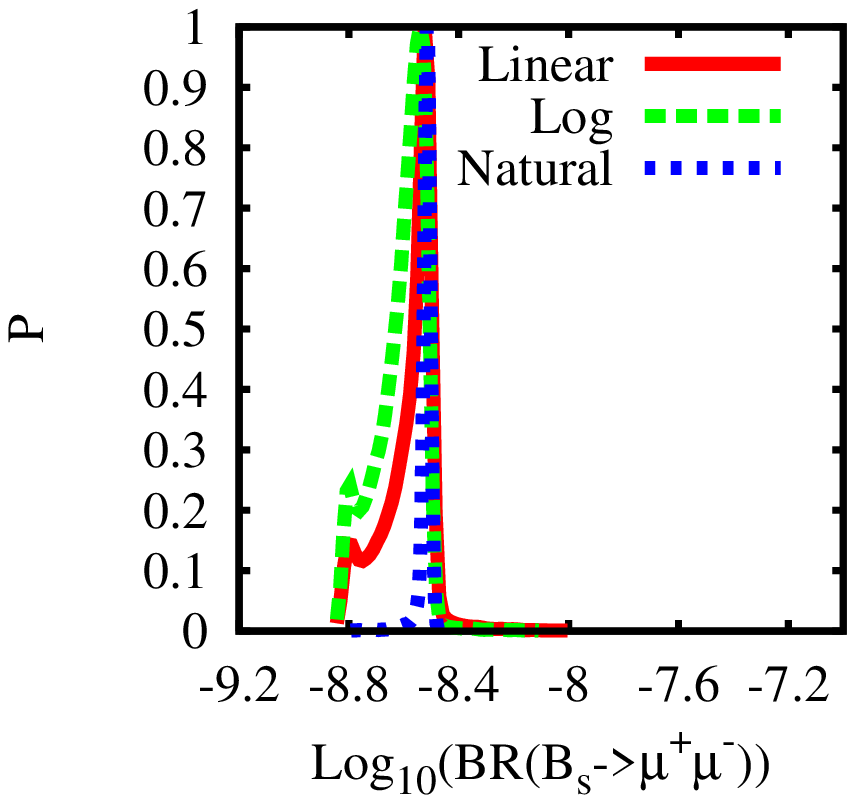}{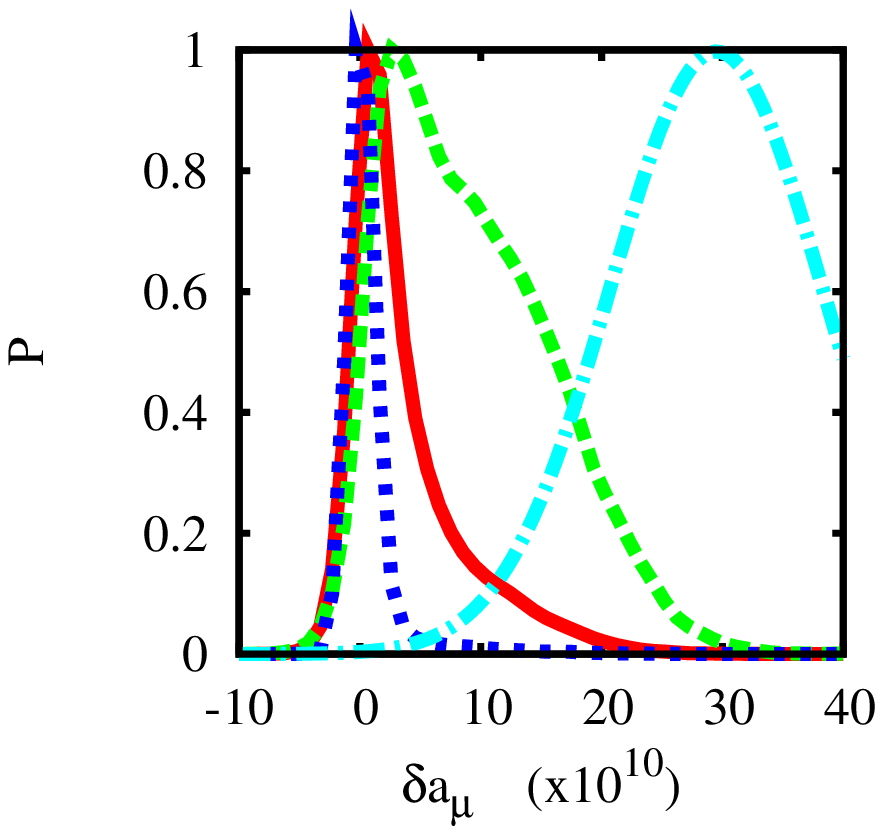}{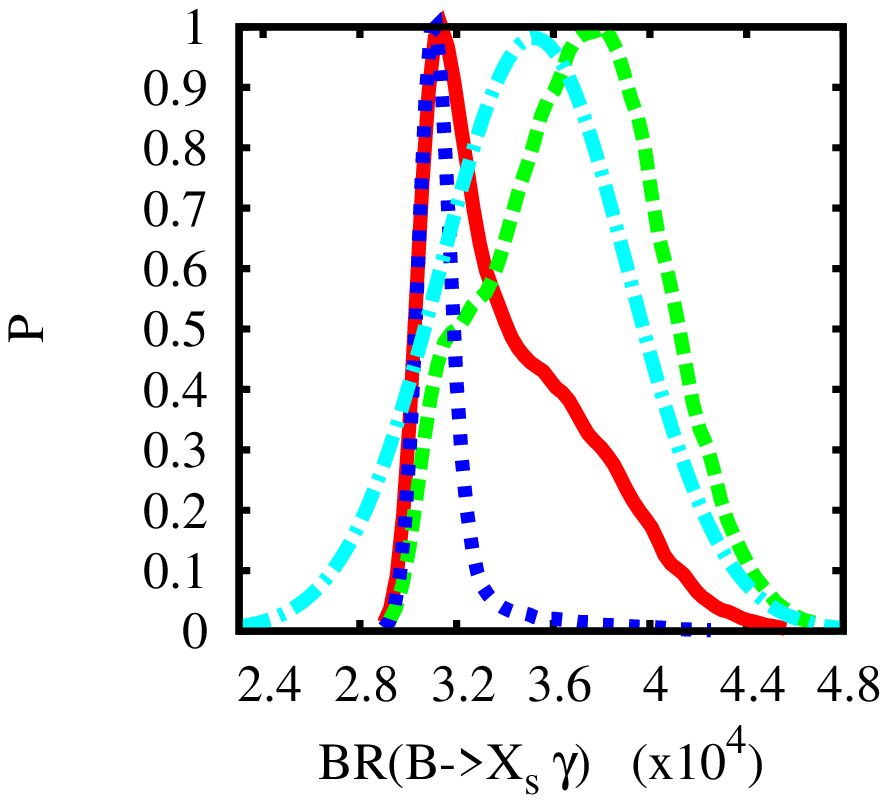}{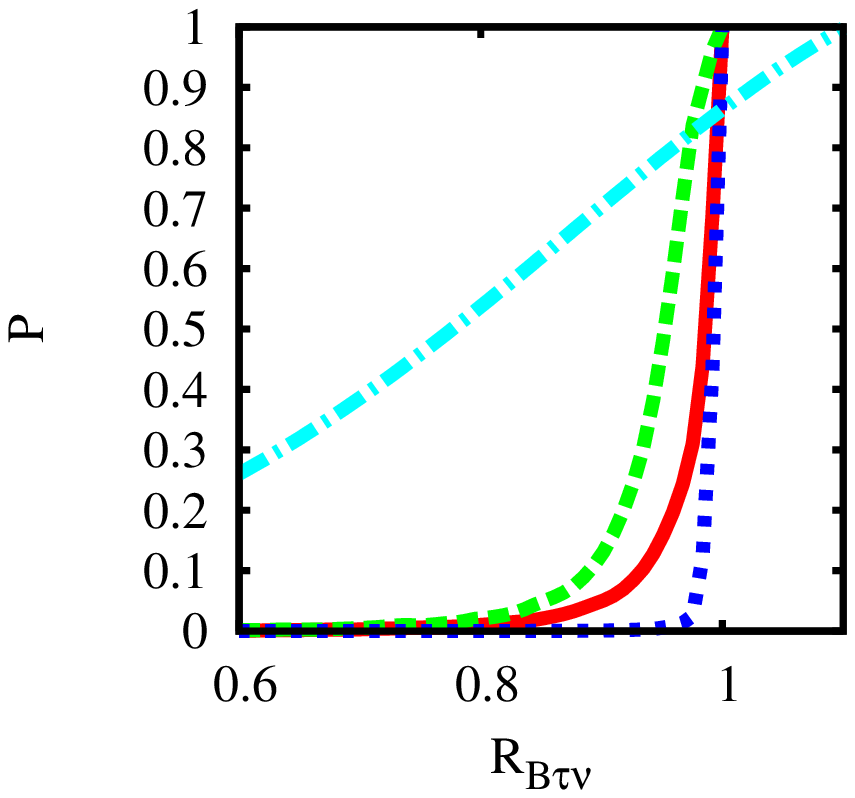}{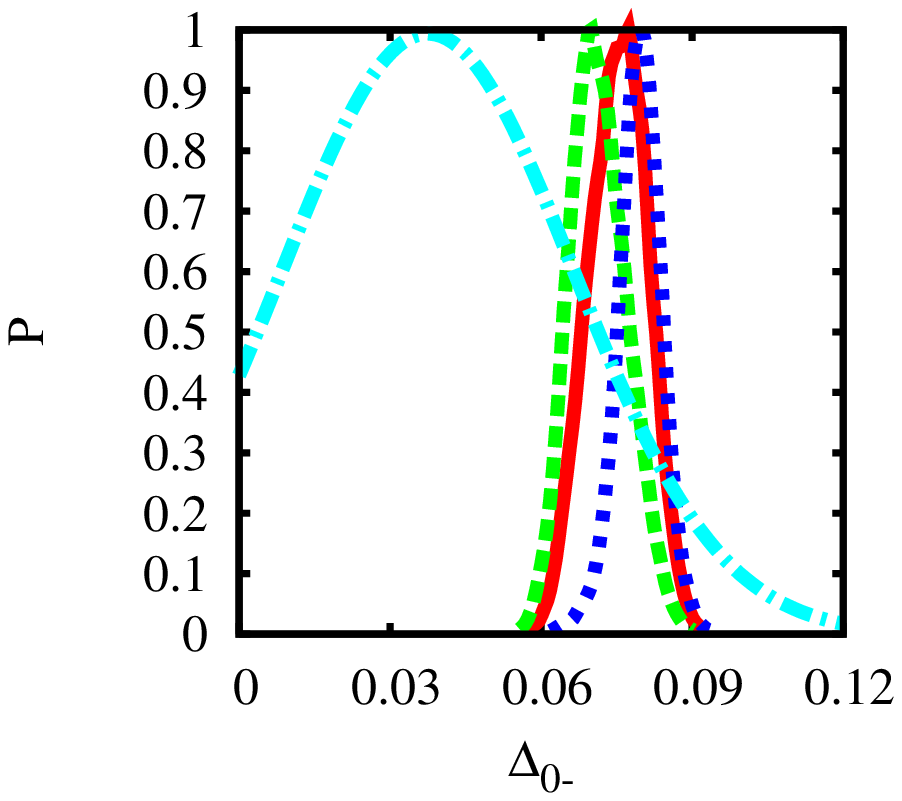}
\caption{The statistical pull of some important constraints on
  mAMSB when the asymmetric dark matter constraint is applied. Posterior PDFs are plotted for: 
red (solid) lines have linear priors,
green (dashed) lines have log priors, blue (dotted) lines have natural
priors and cyan (dash-dotted) is the experimental likelihood
constraint. The figures show (a) the top mass $m_t$, (b) the W mass $m_W$, (c)
the logarithm of the branching ratio for $B_s\to \mu^+ \mu^-$ with the black
arrow being the current  
95$\%$ CL experimental upper limit, (d) the anomalous magnetic moment
$\delta a_{\mu}$, (e) $BR(B\to X_s\gamma)$, (f) the branching ratio 
$BR(B_u \rightarrow \tau\nu)$ divided by its SM prediction and (g) the isospin
asymmetry in $B \rightarrow K^* \gamma$ decays $\Delta_{0-}$.} 
\label{fig:AMSBpull}
\end{center}
\end{figure}

Fig.~\ref{fig:AMSBpull} shows the 1D posterior PDFs for some of the
observables which constrain the mAMSB likelihood for the asymmetric $\mathcal{L}_{DM}$. A comparison with
Fig.~\ref{fig:GMSBpull} shows that many of the posteriors  of the observables
are similar in the mAMSB and mGMSB cases. Only $\delta a_\mu$ and $BR(B \to
X_s   \gamma)$ show an large difference. 
The posterior PDF
for $m_t$ in  Fig.~\ref{fig:AMSBpull}(a) shows that, similarly to mGMSB, 
mAMSB is also in approximate 
agreement with the experimental constraint. Fig.~\ref{fig:AMSBpull}(b) shows
the posterior PDF for $m_W$ which is 
under-predicted by mAMSB similar to the other models we have considered.  
Fig.~\ref{fig:AMSBpull}(c) shows the logarithm of the branching ratio for the
process $B_s\to \mu^+\mu^-$.
The central SM prediction lies at the peak of each posterior.
Our prediction neglects
flavour mixing in the squark sector which result from the AMSB
soft-breaking terms~\cite{arXiv:0902.4880}. 
Ref.~\cite{arXiv:0902.4880} has shown that the effects of including squark
flavour mixing can decrease 
$BR(B_s \rightarrow \mu^+ \mu^-)$ by up to a factor of two for $\tan \beta>22$. 
This would result in a shift of $\log_{10}(BR(B_s \rightarrow \mu^+ \mu^-)$ in
the plot by -0.3. 
The 95$\%$ upper experimental limit
placed on this process is 
$<5.8\times10^{-8}$~\cite{arXiv:0712.1708}, which is always comfortably evaded even
without this additional negative correction. 

The SUSY contribution to the anomalous magnetic moment of the muon is presented in
Fig.~\ref{fig:AMSBpull}(d). With natural priors, the SUSY
contribution to the anomalous magnetic moment
is very small due to heavy sparticles
and low value of $\tan\beta$. Light sparticles and intermediate values of
$\tan\beta$ favoured by the log priors allow a better fit to experiment. 
Larger values of $\tan\beta$, while being consistent with $\delta a_{\mu}$,
would then lead to a worse fit to $BR(B\to X_s \gamma)$. 

Squark flavour mixing corrections, which we neglected, can in principle
have an effect on the $BR(B \to X_s \gamma)$ prediction in mAMSB, but the
effect is 
small~\cite{arXiv:0902.4880}.  
The $BR(B\to X_s \gamma)$ posterior PDF shown in Fig.~\ref{fig:AMSBpull}(e)
approximately follows the experimental constraint for 
log priors, in contrast to mGMSB\@. The natural priors case follows the
experimental constraint less 
closely and the linear priors somewhere
in between the natural and log prior cases. This behaviour is consistent with
the values of the evidence in Table~\ref{tab:model-prob-odds}, where the log priors allow a better fit than the
other two priors.

The two observables shown in Fig.~\ref{fig:AMSBpull}(f,g) are the ratio
$R_{B\tau\nu}$ and the isospin asymmetry in $B$ meson decays $\Delta_{0-}$,
respectively. Both of these observables show the same behaviour as in mGMSB:
not very constraining, 
due
to the weakness of the large uncertainty in the SM prediction. 
While there is a small disagreement between
the mAMSB prediction of $\Delta_{0-}$ and the experimental constraint there is
too little variation of this observable over parameter space for it to mould
the posterior significantly. The behaviour of the electroweak
observables is qualitatively the same as in mGMSB, with all parameters
except $A_{fb}^{0,b}$ and $A_{LR}^0$ being in good agreement with the data while
showing little variation over parameter space. As the accuracy with which
these parameters are known will not improve in the near future, we are
skeptical that they will prove useful in constraining the parameter space of 
simple models such as mAMSB or mGMSB\@. 

\subsection{Combined $\delta a_{\mu}$--$BR(B \to  X_s \gamma)$ constraint}\label{sec.discr}

It is interesting that mAMSB is the only
model we study for which the SUSY contributions to
$(g-2)_{\mu}$ and $BR(B\to X_s \gamma)$ are both probably of the right sign and
magnitude to satisfy 
the experimental constraints simultaneously. There are a number of factors which contribute to this.

While the experimental and SM predictions for $BR(B\to X_s \gamma)$ are consistent with each other, the fact that the experimental result is larger than that predicted by the SM means that given a SUSY contribution to this observable of a specific magnitude, a positive contribution will always agree with the data better than a negative one. mAMSB with $\mu>0$ can accommodate such a contribution  due to the sign of the coupling $A_t$. In mGMSB, mSUGRA and the LVS this will not be the case in the majority of parameter space. 
However, we note that using a more recent evaluation~\cite{gambino} of the Standard Model prediction which gives the branching ratio as $3.28\pm 0.25 \times 10^{-4}$ would lessen this effect somewhat.

A further effect is the hierarchies
present in mAMSB spectra compared to the relatively compressed mGMSB spectra. 
Recall from Eq.~\ref{eq:g-2} that the magnitude of the
supersymmetric one-loop $g-2$ correction scales as $\sim M_{1,2} \mu \tan
\beta/m^4$, where $m$ is the mass scale of the relevant particles in the loop, 
i.e.\ the chargino/sneutrino or neutralino/smuon mass scale. 
The dominant SUSY contributions to  $BR(B\to X_s \gamma)$ come from
top-charged Higgs and stop-chargino contributions. Agreement with the data
typically prefers that the SUSY contribution to the branching ratio is not too
large, i.e.\ the charged Higgs and stop
masses are not too small. Thus, models in which the ratio of these masses to 
the slepton/neutralino masses is large have the potential to fit both
$(g-2)_\mu$ and $BR(B\to X_s \gamma)$ simultaneously. 
In mGMSB, the spectrum is rather compressed compared to mAMSB, and so models
with light neutralinos and sleptons also tend to have relatively light charged
Higgs and stops. Thus, mGMSB fits the combination of the two observables less
well than mAMSB.
The interplay between $(g-2)_\mu$ and  $BR(B\to X_s \gamma)$ fits has
recently been explored in detail in~\cite{Feroz:2009dv} for the case of the
CMSSM, and previously also in~\cite{belyaev1,belyaev2}.

\begin{figure}
\begin{center}
\twographsB{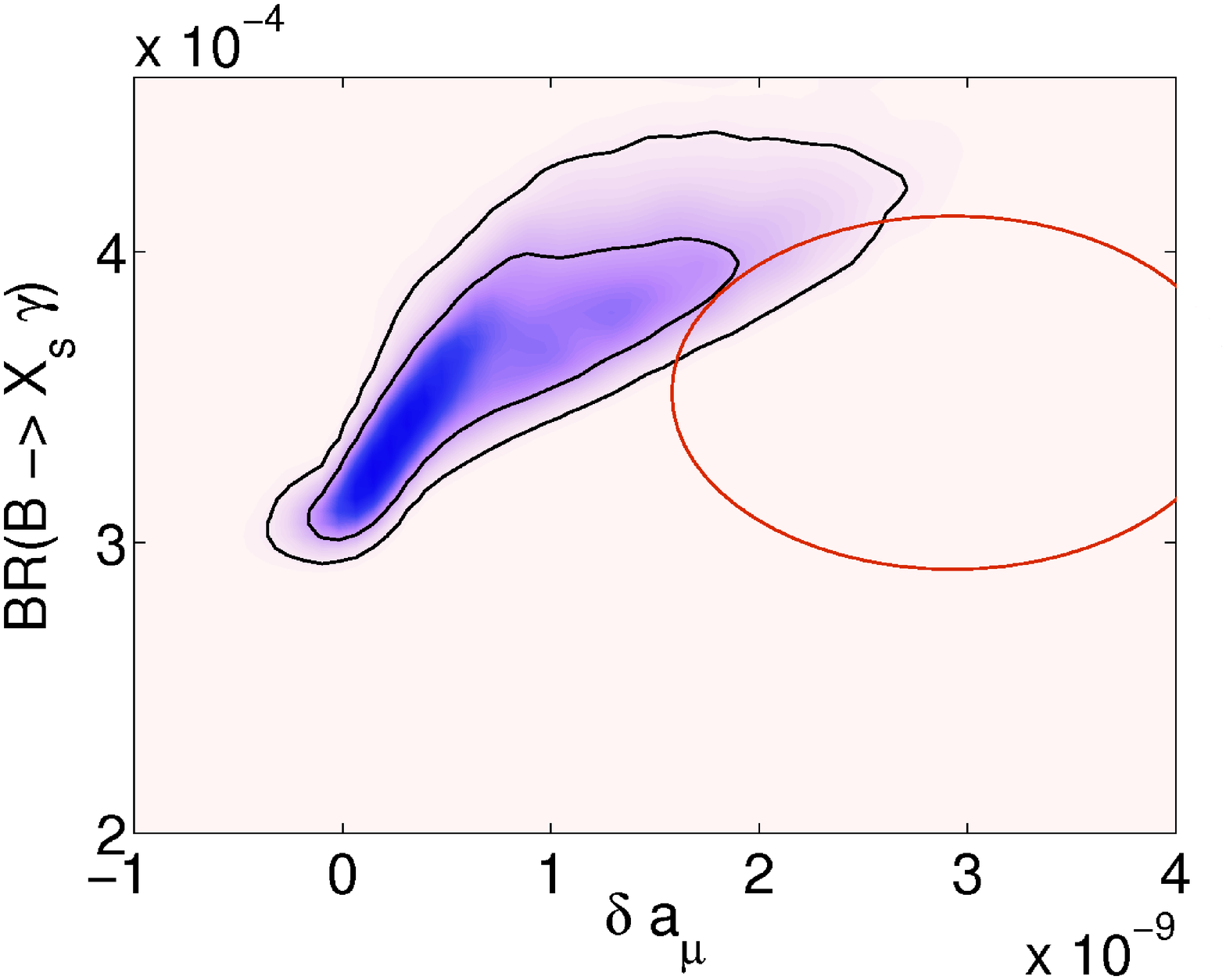}{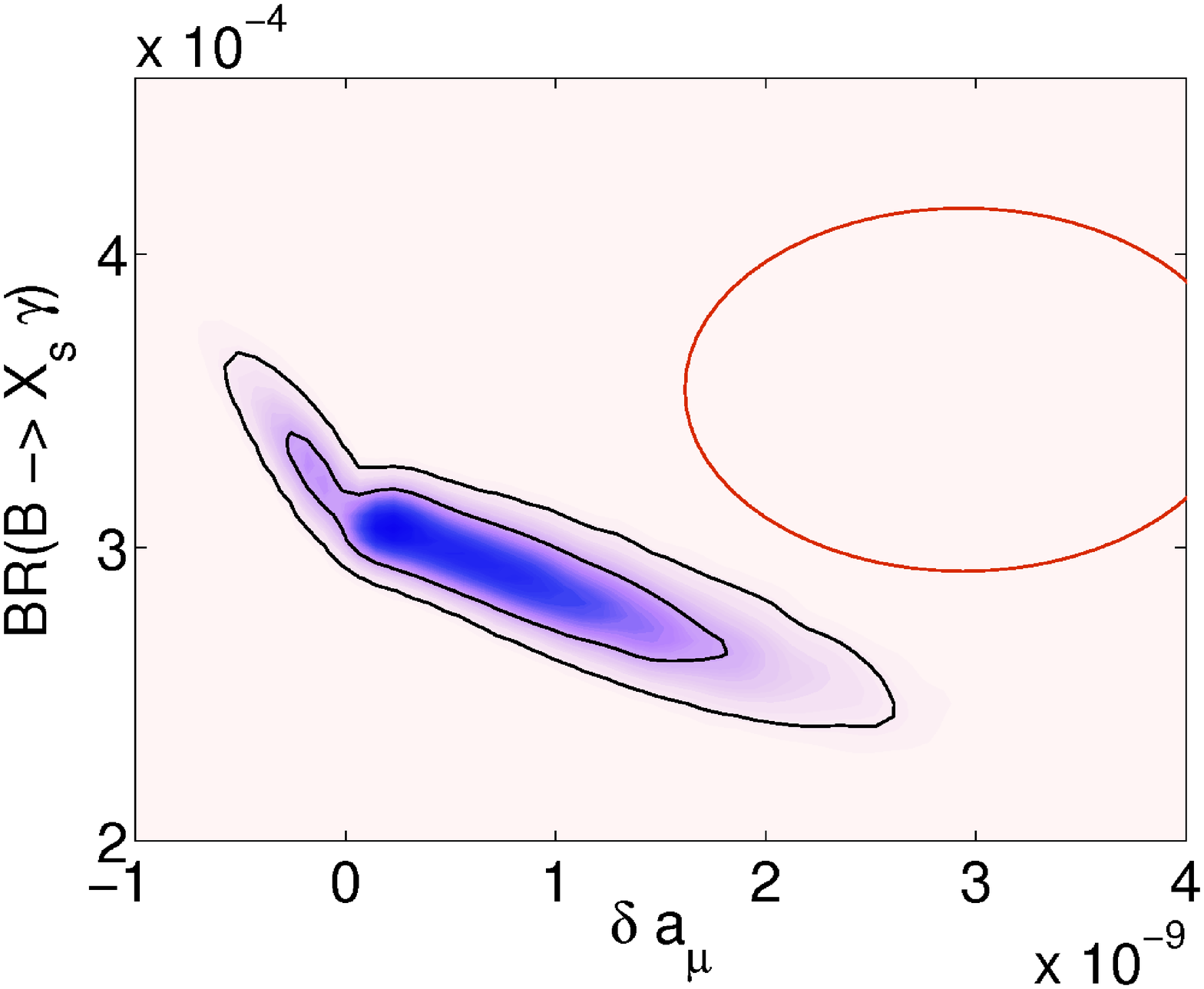}
\caption{The 2D joint posterior in the $(g-2)_{\mu}$-$BR(B\to X_s\gamma)$ plane
  for (a) mAMSB and (b) mGMSB\@. Log priors and no dark matter constraint apply in both cases. 68$\%$ and
  95$\%$ Bayesian credibility regions are shown as the inner and outer curves,
respectively. The ellipse is centered on the experimental values of the
two observables and show the 68$\%$ confidence level bounds on $\delta
a_{\mu}$ and $BR(B\to X_s \gamma)$.} 
\label{fig:g-2bsg}
\end{center}
\end{figure}
 We illustrate this behaviour in
Fig.~\ref{fig:g-2bsg} which shows the 2D posterior PDFs in the
$(g-2)_{\mu}$-$BR(B \to X_s \gamma)$ plane for mAMSB and mGMSB with log
priors and no dark matter constraint. It shows the correlation between
the two observables for 
mAMSB and anti-correlation for mGMSB, along with red ellipses 
showing the combined experimental constraint. This (anti-)correlation is
expected from the signs of $A_t$ and $M_3$ in the models~\cite{arXiv:0902.4880}.
Parts of mAMSB parameter space that provide a good fit to all of the data 
are also within the error ellipse, contrary to the mGMSB case.

\subsection{Best-fit points\label{sec:bestfit}}

\begin{figure}[!ht] 
\begin{center}
  \unitlength=1in
\begin{picture}(6,7)(0,0.5)
\put(0,4.3){\includegraphics[angle=0, width=6in]{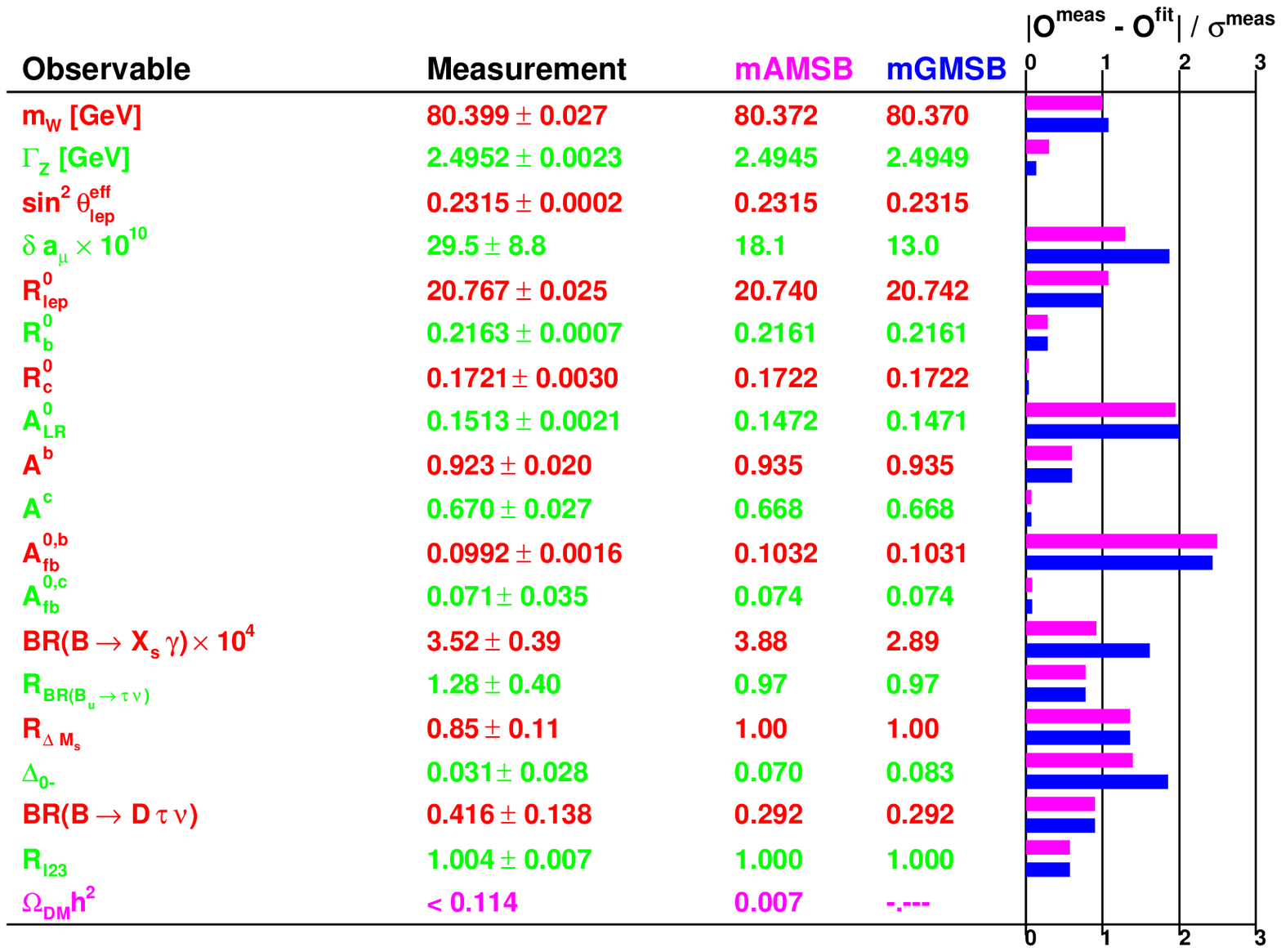} }
\put(0,0){\includegraphics[angle=0, width=6in]{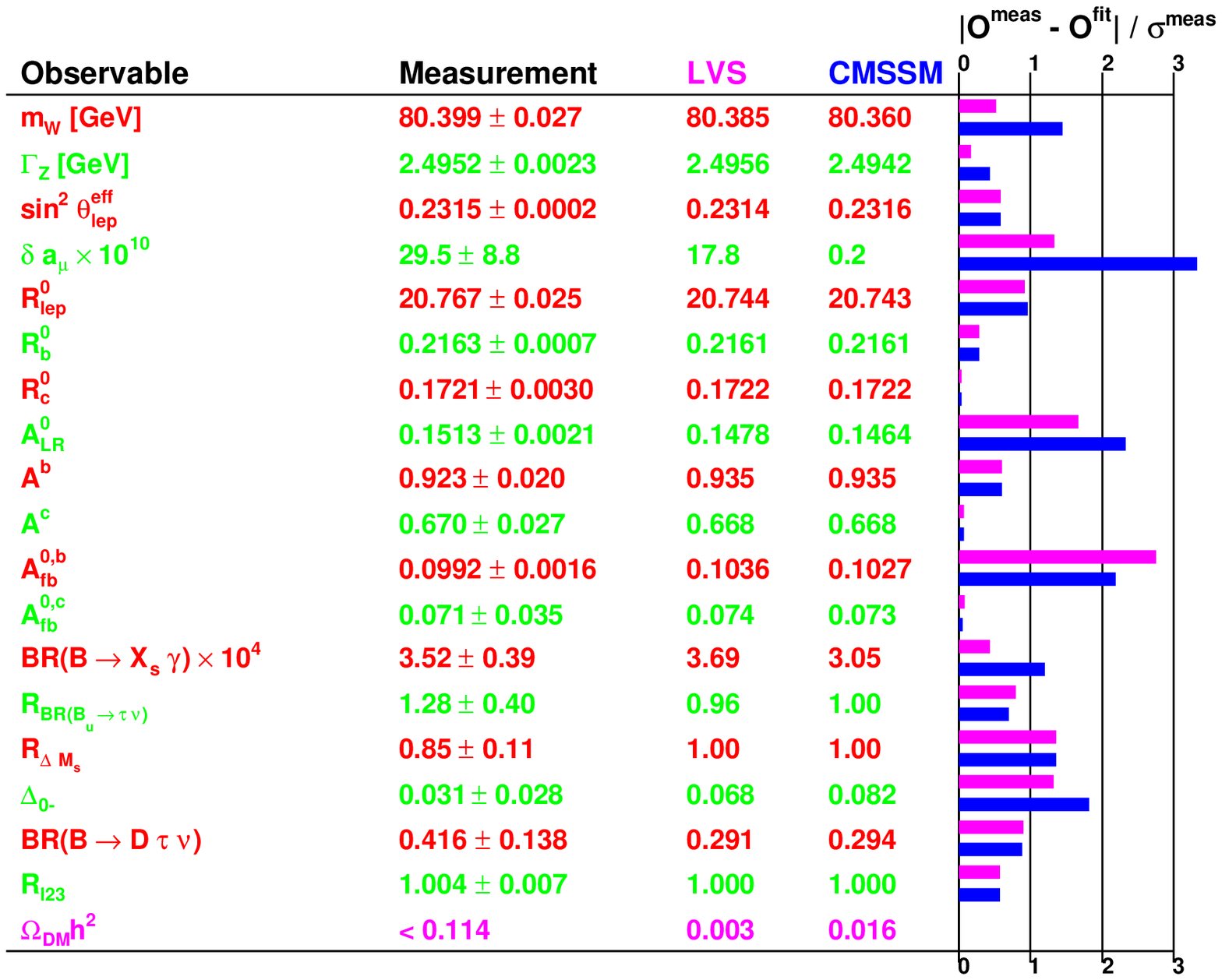}}
\end{picture}
\end{center}
 \caption{Best fit points.
   The best-fit point prediction and experimentally measured values for each
   observable are listed. The pulls, represented by the horizontal bars,
   are in units of standard deviations. The corresponding parameter points
   are: $m_0 = 312.7$ GeV, $m_{3/2} = 45$ TeV, $\tan
   \beta = 15.9$ for mAMSB, $N_{mess} = 7$, $M_{mess} = 1.23 \times
   10^{7}$ GeV,$\Lambda=19.5$TeV $\tan \beta = 15.9$ for mGMSB, $m_{1/2} = 328$ GeV, $\tan \beta = 11.6$ for LVS and $m_0 =
  3338$ GeV, $m_{1/2} = 382$ GeV, $A_0 = 635$, $\tan \beta = 8.6$
  for the CMSSM. $\mu$ is positive for all of the best-fit points.}  
 \label{fig.ewwg1} 
\end{figure}
Fig.~\ref{fig.ewwg1} shows the best-fit points (defined to be the highest
likelihood point) for mAMSB, mGMSB, the CMSSM and LVS\@. The asymmetric
DM constraint was taken, except for mGMSB where no DM constraint was
applied.
The caption contains the parameters of each model that yielded the best-fit,
and the bars on the right-hand side measure how far each observable is from
its experimental central value. 
We should note that our fits contain quasi-degenerate likelihood maxima, and 
this combined with the fact that the {\sc MultiNest} algorithm is not
optimized to find the best-fit point, may mean that the parameters of the
best-fit point are poorly determined. This is not important for any of the
Bayesian inferences we have presented, but it is instructive to examine the
properties of some point in each model. The point sampled with the highest
likelihood is a good candidate for the point to choose.
The CMSSM best-fit point is in the focus
point region~\cite{Feng:1999zg}. It is not a really good fit to the
anomalous magnetic moment of the muon, since heavy sleptons and moderate $\tan
\beta$ render the SUSY contribution small. The other models' best-fit points
all fit $\delta a_\mu$ much better, owing to their lighter spectra. Comparing
the pulls of the electroweak 
observables, we see that there is not much variation in the $\chi^2$ values
coming from them across the models except for $M_W$. The forward-backward
asymmetry of $e^+e^- \rightarrow b \bar b$, $A^{0,b}_{fb}$, shows a small 
change, being a little larger in the LVS than the other models. The pull of
the left-right asymmetry $A_{LR}^0$ also shows some small dependence. 
All of the models listed have a predicted relic density of dark matter much
smaller than the WMAP-inferred central value, requiring a dominant
component of non neutralino dark matter (except for mGMSB, where we do not
apply the DM constraint).

\section{Conclusion}

We ask the question: are any of the most commonly assumed low-parameter
SUSY breaking 
mediation mechanisms favored over the others by current indirect and
cosmological data? Aside from direct searches, SUSY corrections contribute to
the observables we consider only in loop effects. For observables that agree
with 
the standard model prediction, arbitrarily heavy sparticles suppress such
loop effects and so will fit that observable. However, the anomalous magnetic 
moment of the muon prefers a significant contribution from supersymmetric
loops, and so this observable 
prefers light SUSY particles. The dark matter constraint from cosmological
observations cannot be accounted for by the Standard Model, but can be fitted
by the MSSM if the lightest supersymmetric particle is the neutralino. 
We argue that the Bayesian 
evidence is the relevant statistical quantity for such an analysis.

When performing such a statistical global fit, it is important to check
robustness. The fits should be dominated by the data and not by the form of the
prior if we are to claim robustness. Previous studies showed that fits to the
CMSSM with four extra parameters are prior
dependent~\cite{Allanach:2006jc,hep-ph/0609295,Trotta:2008bp}, whereas the LVS 
model~\cite{Allanach:2008tu} 
(with only two extra parameters) is more robust. Thus it was natural to select
mGMSB and mAMSB, each of which have three extra parameters to check robustness
of the fits and compare the models against each other. 
One expects the level of robustness to go down with higher numbers of
parameters, which a recent fit to the phenomenological
MSSM~\cite{AbdusSalam:2009qd} with 
twenty extra parameters,  illustrated\footnote{There were a couple of
  prior independent inferences in the fits, such as the lightest CP-even Higgs
mass.}.
We have presented constraints on the mGMSB and mAMSB
parameter spaces, and found that in both cases there {\em is}\/ significant prior
dependence. Parameter inference from the models are therefore not robust and
therefore require further more precise and direct data, perhaps from
collider measurements of SUSY particles. 
Unsurprisingly, no robust statement regarding the sign of $\mu$ can be made for
mAMSB or mGMSB\@. We
therefore encourage that 
future work on mAMSB and mGMSB include both signs of $\mu$ until the data is
strong enough 
to support a prior-independent pronouncement on the status of the sign of
$\mu$, and not to disregard the $\mu<0$ branch of the theory based on the
preference of one observable for $\mu>0$.
The Large Volume 
Scenario, with two fewer parameters than the CMSSM is more constrained and shows
a robust moderate preference for $\mu>0$.

The model preferred by the data depends on what we assume for the DM relic
density: whether it is  
made entirely of neutralinos (symmetric constraint) or whether we allow for
the presence of non-neutralino dark matter (asymmetric constraint). 
An analysis of the
constraining power of the various observables showed that it
resides dominantly in the DM constraint in the case of the CMSSM and the LVS. This is not the case in mAMSB
where the relic density is uniformly too small by an order of magnitude across parameter space, and the main constraint comes from the combined electroweak observables. 
Dropping the DM constraint altogether allows a comparison with mGMSB, but then
no strong robust preference for any model can be found. 
However, for the symmetric constraint, mAMSB is strongly disfavoured (since it
predicts 
essentially no neutralino dark matter) over the CMSSM and LVS. 
With the asymmetric constraint, mAMSB is at least moderately favoured over
the CMSSM. 

Experience and
familiarity with the methods of model selection and Bayesian inference 
from work such as that contained here will be
invaluable once further more constraining data become available, hopefully
from SUSY signals at colliders.  

\section*{Acknowledgements}
MJD wishes to thank St. John's College, the CET and EPSRC for
support. SSA is supported by the Gates Cambridge Trust, and thanks F.~Quevedo and D.~J.~C.~McKay for discussions.  We thank A.~Casas and C.~Lester for discussions about the MSSM prior and P.~Slavich for
communication about $BR(B \rightarrow X_s \gamma)$.  This work was performed using the Darwin Supercomputer of the University of Cambridge High Performance Computing Service (http://www.hpc.cam.ac.uk/), provided by Dell Inc.\ using Strategic Research Infrastructure Funding from the Higher Education Funding Council for England. The original idea for this
project came from discussions with S.~Kraml and others in the TeV Colliders Les
Houches workshop in 2007. 
This work was partially supported by STFC. 

\appendix
\section{Dark Matter Direct Detection \label{sec:DM}}

In this appendix we present results on DM direct detection cross-sections for
the Large 
Volume Scenario. Direct detection rates in the 
CMSSM have been most recently been presented in Bayesian global fits in
Refs.~\cite{arXiv:0705.2012, arXiv:0806.1923}. Our 
updated CMSSM fits are similar to
the results of those articles, so we do not include them here. 
Our calculations of the detection cross-section have been obtained with
\texttt{micrOMEGAS2.3.1}\cite{Belanger:2008sj,Belanger:2006is,Belanger:2004yn,Belanger:2001fz}. 

Neutralinos are Majorana fermions and can therefore interact with quarks via
scalar and axial-vector  
interactions, but not via a vector-like interaction. In the scalar (spin-independent) case, the neutralino can 
interact at tree-level with quarks in the nucleus via Higgs or squark exchange. At one-loop the neutralino 
can couple to gluons via a quark/squark loop. At zero momentum transfer, the total cross-section is given by
\begin{equation}
 \sigma = \frac{4m_{red}^2}{\pi} \left( Zf_p + (A-Z)f_n \right)^2
\end{equation}
where $m_{red}$ is the nucleon-neutralino reduced mass, $Z$ and $(A-Z)$ are the number of protons and neutrons in 
the nucleus and $f_{p(n)}$ is the coupling of the neutralino to the proton(neutron) respectively.

The neutralino-quark axial-vector interaction leads to a spin-dependent coupling proportional to $J(J+1)$, where
$J$ is the spin of the nucleus. The cross-section is given by 
\begin{equation}
 \frac{d\sigma}{d|\vec{v}|^2} = \frac{1}{2\pi v^2} \overline{|T(v^2)|^2}
\end{equation}
where $v$ is the relative velocity of the neutralino and $T(v^2)$ is the scattering matrix element. This
expression is then integrated over the Boltzmann velocity distribution of the neutralinos to obtain an average
cross section.

Since the results of our fits indicate that the LVS is sufficiently constrained so as to be independent of the
prior, we show only the case of linear priors. The
cross-sections are independent of whether the nucleon is a proton or
neutron. We therefore have plotted the 
average $(\sigma^p+\sigma^n)/2$ for each point sampled. As elsewhere in this
article we marginalize over the sign 
of $\mu$. 

\begin{figure*}
\begin{center}
\twographsC{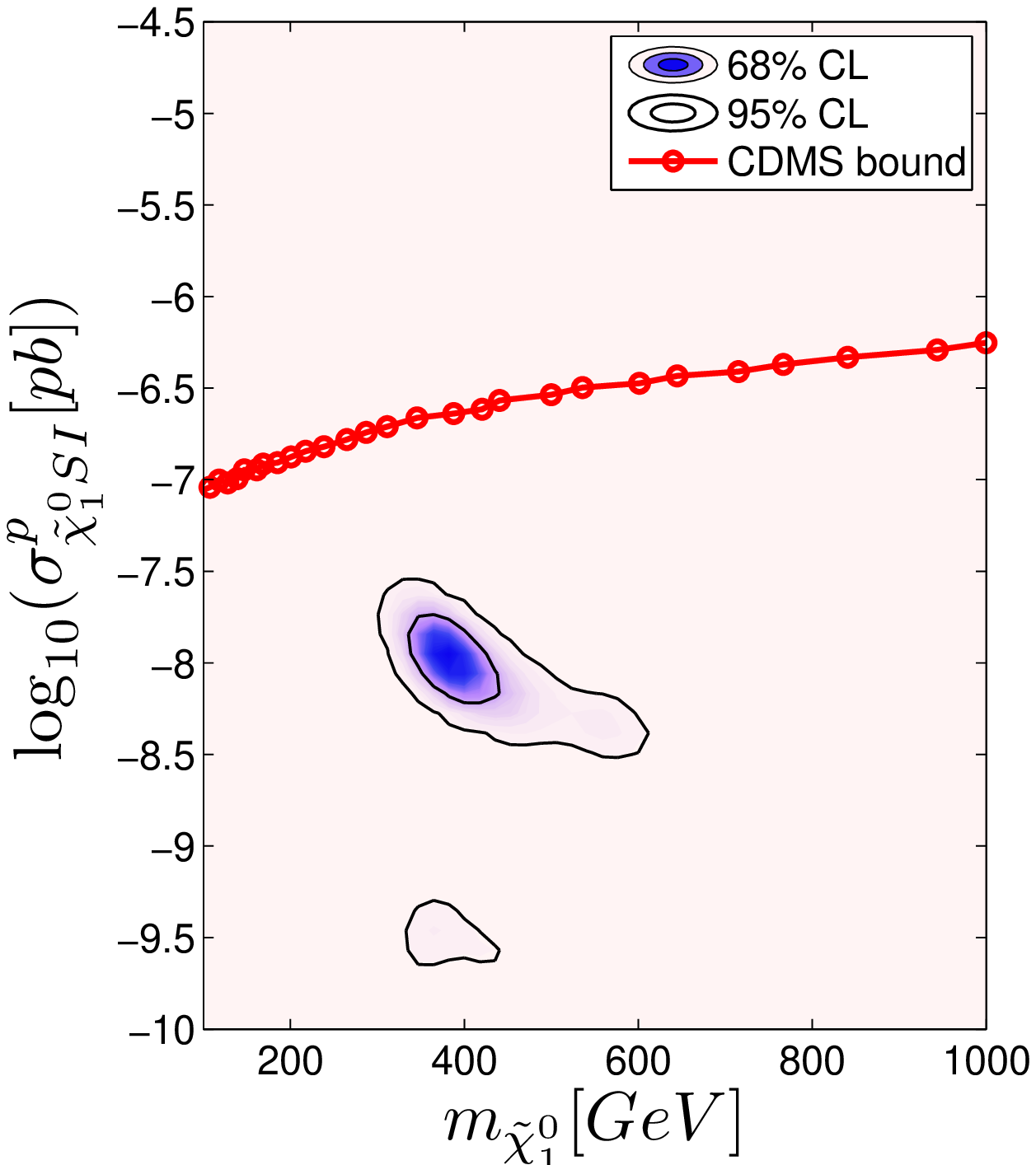}{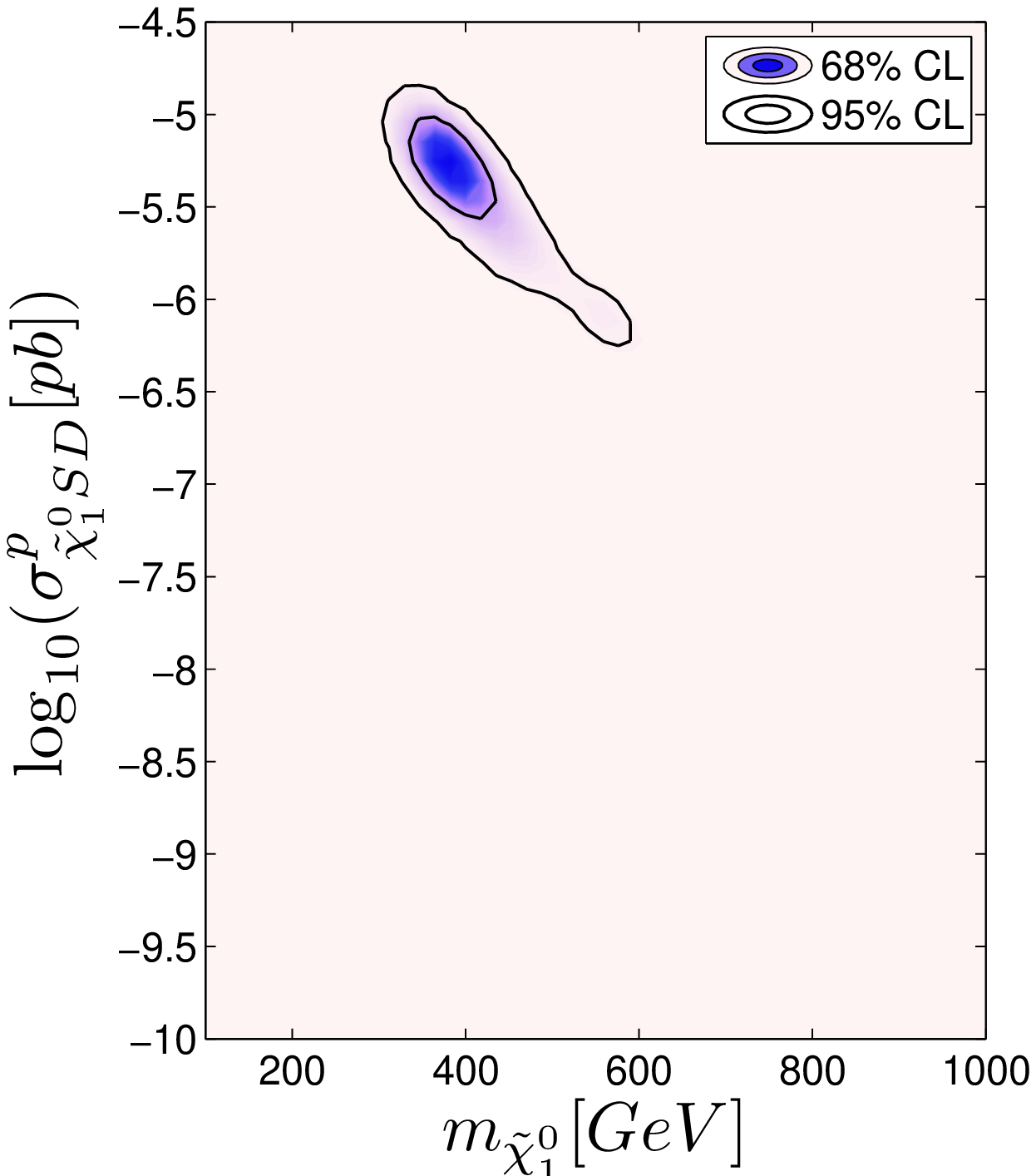}
\caption{Posterior PDFs with linear priors for (a) the spin-independent cross-section $\sigma_{SI}$ and (b) the spin-dependent cross-section $\sigma_{SD}$ for neutralino-nucleon elastic scattering plotted against the neutralino mass $m_\chi$ in the Large Volume Scenario, marginalized over both signs of $\mu$. Both plots have a symmetric dark matter constraint.}
\label{fig:DMdetection}
\end{center}
\end{figure*}

Fig.~\ref{fig:DMdetection}(a) shows the 2D posterior PDF for the logarithm of the spin-independent cross-section per nucleon in
picobarns against the neutralino mass $m_{\chi}$.
We also show the CDMS~\cite{arXiv:0802.3530} upper bound assuming the local DM density to
be $\rho_0 = 0.3\mbox{Gev/cm}^3$. 
Fig.~\ref{fig:DMdetection}(b) shows the posterior PDF for the log of the
spin-dependent cross-section per nucleon versus the neutralino mass. The DM
detection cross-sections are both well below the bounds
set by the current generation of DM searches such as CDMS\cite{arXiv:0802.3530} in the
spin-independent and 
SuperK~\cite{hep-ex/0404025} in the spin-dependent cases. Future one-tonne detectors
should be able to probe the entire 
range of the LVS parameter space for the spin-independent cross-section down
to $\sigma_{SI}\sim 10^{-10}$pb. 
The spin-dependent result is independent of the sign of $\mu$ giving a robust
prediction of $\sigma_{SD} \sim 
10^{-5.5\pm0.5}$pb.


\providecommand{\href}[2]{#2}\begingroup\raggedright\endgroup

\end{document}